\documentclass{article}
\usepackage{amsmath, amssymb, amsfonts}
\usepackage{bm}
\usepackage{dsfont} 
\usepackage{graphicx}
\usepackage{subcaption}
\usepackage{pgfplots}
\usepackage{booktabs,xcolor,siunitx}
\usepackage{caption}
\usepackage{subcaption} 
\usepackage{hyperref}
\definecolor{DarkBlue}{rgb}{0.00,0.00,0.55}
\definecolor{Black}{rgb}{0.00,0.00,0.00}
\hypersetup{
linkcolor = DarkBlue,
anchorcolor = DarkBlue,
citecolor = DarkBlue,
filecolor = DarkBlue,
colorlinks = true,
urlcolor = Black
}
\usepackage{geometry}
\usepackage{hyperref}
\usepackage{xcolor}
\DeclareMathOperator*{\argmin}{arg\,min} 
\DeclareMathOperator*{\argmax}{arg\,max} 

\usepackage{algorithm}
\usepackage{algpseudocode}
\usepackage{amsmath} 
\usepackage{amssymb} 
\usepackage{amsfonts} 
\usepackage{tabularx} 
\usepackage{authblk} 

\title{A copula-based variational autoencoder for uncertainty quantification in inverse problems: Application to damage identification in an offshore wind turbine}
\author[1]{Ana Fernandez-Navamuel}
\author[2]{Martin A. Díaz Viera}
\author[1]{Matteo Croci}

\affil[1]{Basque Center for Applied Mathematics (BCAM), Bilbao, Spain  Bilbao, Spain}
\affil[2]{Instituto Mexicano del Petróleo, Eje Central Lázaro Cárdenas Norte 152, San Bartolo Atepehuacán, Gustavo A. Madero, C.P. 07730, Ciudad de México, Mexico}
\date{\today}

\begin{document}

\maketitle
\begin{abstract}
    Structural health monitoring of floating offshore wind Turbines (FOWTs) is critical for ensuring operational safety and efficiency. However, identifying damage in components like mooring systems from limited sensor data poses a challenging inverse problem, often characterized by multiple solutions where various damage states could explain the observed response. To overcome this, we propose a variational autoencoder (VAE) architecture, where the encoder approximates the inverse operator that maps the observed response to the system's condition, while the decoder approximates the forward operator that maps the system's condition and measured excitation to its response.
    In the proposed methodology, the observed response corresponds to some representative features derived from short-term rotation motion signals of the FOWT platform. The damage condition is described by the severity level of two damage classes frequently found in the mooring system (anchoring and biofouling). 
    This work proposes a novel copula-based VAE architecture that decouples the marginal distribution of variables from their dependence structure, thus enabling the representation of complex posterior distributions.
    We provide a comprehensive comparison of the copula approximation against standard Gaussian and Gaussian mixture approaches.
    Numerical experimentation, conducted on a high-fidelity synthetic dataset, demonstrates that the Gaussian Copula VAE provides a more scalable alternative to Gaussian mixtures in high dimensions. 
    Indeed, the copula achieves the same performance with significantly fewer parameters than the Gaussian Mixture alternatives, whose parametrization grows prohibitively with the dimensionality of the latent space. These results suggest that copula-based VAEs provide a robust framework for uncertainty-aware damage detection in FOWT mooring systems.
\end{abstract}

\section{Introduction}
Continuously assessing the health condition of floating offshore wind turbine (FOWT) mooring systems is crucial to ensure their durability and thus increase their contribution to the power network. 
This paper proposes an efficient copula-based variational autoencoder (VAE) for identifying damage in the FOWT mooring system using limited motion responses measured during monitoring.
Once trained, the VAE is able to process new motion observations in real time and provide a probabilistic description of the most likely underlying FOWT health conditions. 
The copula VAE thus identifies multiple solutions to the inverse problem and provides a measure of the uncertainty of the FOWT diagnostic.

Offshore wind energy solutions offer outstanding energy production opportunities owing to the strong winds that blow in the deep-water sites where they are installed~\cite{Gibbs03102022}, and have thus become increasingly attractive for the European green transition~\cite{gonzalez2025offshore}. 
Indeed, 117 GW of new wind power capacity was installed globally in 2024 only~\cite{gwec2025}. 
However, the appeal of building wind turbines offshore is counterbalanced by a few drawbacks: harsh operating conditions, reduced accessibility~\cite{JAHANI2022111136}, and the technical complexity of installing the required mooring and floating systems~\cite{xu2021design}.
These challenges result in expensive and potentially hazardous maintenance operations, which may expose workers to hazardous environments during inspection and repair interventions~\cite{Milligan_Workers1014}. 

Mooring systems are one of the most critical components in floating offshore wind turbines (FOWTs). 
These systems ensure an adequate anchorage to the seabed and limit the platform's movements in the presence of deep currents, strong wind speeds, and waves. 
The harsh, corrosive seawater environment in which FOWTs operate makes mooring systems highly vulnerable to multiple sources of damage. Among these, the most frequent are \emph{biofouling}, caused by the adherence of algae or other organisms~\cite{spraul2017effect}, and \emph{anchoring}, i.e., the trawling of the mooring system anchor~\cite{LIU2021391}. 
If these phenomena remain undetected over a prolonged period, they may cause the platform to move excessively, thereby compromising the FOWT's integrity and its power production capability. 

In this context, structural health monitoring (SHM) is vital for continuously and remotely assessing the current state of FOWTs' components. 
Its goal is to infer the condition of the mooring system from a limited set of sensors that measure its response, and can thus be posed as an inverse problem. 
A wide range of sensors is available, from low-cost, low-maintenance, and easily deployable sensors (e.g., accelerometers on the tower or inclinometers on the platform) to more complex and expensive ones (e.g., direct tension mooring line measurement)~\cite{CORADDU2024}.
Over the years, significant efforts have been made to exploit these data, ranging from statistical pattern recognition tools~\cite{martinez2016structural} to artificial intelligence (AI)~\cite{KHAZAEE20221568, Pezeshki2023_intro}.

Deep neural networks (DNNs)~\cite{goodfellow2016deep} have become extensively employed in the field of FOWT assessment, from standard fully-connected to convolutional, recurrent, or long-short-term-memory (LSTM) approaches~\cite{chung2020detection, xu2022multisensory, sharma2024convolution, MAO2024, mao2023LSTM-SVM}.
Since inverse problems are often ill-posed (they may admit none or multiple solutions that may not be robust to small perturbations in the available data), regularization strategies have been employed to handle them. 
The authors of~\cite{Pardo_loss_inverse} explore an autoencoder (AE) architecture, where the encoder maps the inverse problem and the decoder represents the forward operator, which delivers the reconstructed measurements. This architecture enables the use of the measurement discrepancy (reconstruction error) as a regularizing term in the loss function, ensuring the physical consistency of the inverse problem estimates. 

However, existing DNN approaches for FOWT damage detection are deterministic and thus do not provide any quantification of the uncertainty in the predictions. This is a severe limitation in real-world applications, where multiple uncertainty sources exist, including environmental and operational variability, measurement error, modeling simplifications and assumptions, among others.
Under such circumstances, the estimated deterministic solution may be far from the actual one. 
Additionally, small perturbations in the observed inputs can result in significantly different outcomes. 
These phenomena compromise the reliability of the method and, as a consequence, the associated decision-making in SHM applications~\cite{Navamuel2022supervised, Navamuel2022deepPCA}. 
Hence, probabilistic approaches are needed to fully describe the solution space of the inverse problem~\cite{xia2022bayesian}. 

Variational autoencoders (VAEs) offer a powerful approach to handling uncertainty during inference~\cite{VAE_tutorial}. Unlike standard autoencoders, VAEs incorporate a probabilistic model that represents the posterior distribution given some prior information~\cite{mo2025explainable}.
VAEs have been employed as generative models and for dimensionality reduction and clustering tasks~\cite{graving2020vae, yong2022bayesian, bond2021deep, portillo2020dimensionality} and for solving inverse problems in scientific computing~\cite{prost2023inverse, wu2022inverse, mcaliley2024stochastic, sahlstrom2023utilizing}. 
In the context of SHM, VAEs have been mostly applied for damage-sensitive feature extraction to detect structural changes from measured response data~\cite{coracca2023unsupervised, pollastro2023semi, lin2024structural}.
VAEs have been employed to infer changes in the material properties of multi-degree-of-freedom  (MDOF) structures~\cite{simpson2021machine, bacsa2025structural}, to classify different structural damage types~\cite{huang2025detecting}, and to estimate the uncertainty in long-term wind turbine blade fatigue estimation~\cite{Mylonas2021}. However, all the aforementioned works~\cite{simpson2021machine, bacsa2025structural, huang2025detecting, Mylonas2021} share the same limitation: they employ normal (Gaussian) posterior probability density functions (PDFs) with diagonal covariance, which lack expressivity and prevent an accurate representation of the uncertainty. 

To enhance the expressivity of VAEs, more complex parametric posterior distributions have been explored.
Among these, mixture models have been proven effective~\cite{kviman2023cooperation, nalisnick2016approximate, LIU201943,Rodriguez2023, Pardo_loss_inverse}. 
In a FOWT context, Gaussian mixtures with diagonal covariance have been employed in~\cite{Navamuel_wes2025} to identify two-dimensional mooring system damage conditions.
Although mixtures increase VAE expressivity, selecting the covariance structure of its terms requires balancing an accuracy/cost trade-off: on the one hand, diagonal covariances lead to efficient methods, yet they lack approximation capabilities. 
On the other hand, full covariance models are expressive, yet pose scalability issues in high dimensions due to the large number of mixture parameters required.

Copula models offer an efficient alternative for describing complex multivariate distributions~\cite{tagasovska2019copulas, hernandez2024fast}.
Copula models separate the dependence structure of joint distributions from the marginal PDFs of each individual random variable, allowing each to be modeled and analyzed separately~\cite{nelsen2003properties, joe2014dependence}.
Copula models have been a cornerstone in finance and risk management~\cite{cherubini2011dynamic, mcneil2015quantitative, carbonera2024variational}, hydrology,
actuarial science, econometrics~\cite{frees1998understanding, patton2012, salvadori2007}, geophysics and reservoir characterization\cite{mcaliley2024stochastic}.
Parametric copulas enable the joint stochastic simulation of petrophysical and elastic properties \cite{Vazquez2023} and form the core of sophisticated geostatistical seismic inversion frameworks within a Bayesian setting \cite{Vazquez2025} as an alternative to approaches based on Gaussian mixtures \cite{Grana2017, Figueiredo2019}. Beyond these traditional fields, copulas are now driving innovation in modern computational domains. In image reconstruction, they model local dependencies in mixed data types \cite{suh2016gaussian}, while in language modeling, they enhance generative architectures: Gaussian copula-based VAEs capture latent dependencies \cite{wang2019neural}, variational methods use the copula posterior to preserve dependency structures \cite{zhong2021variational}, and contrastive copula frameworks isolate spurious couplings for improved classification \cite{wu2023c}. 
Overall, from foundational risk analysis to cutting-edge generative AI, copulas have established themselves as a cross-disciplinary and indispensable statistical tool for precisely modeling multivariate dependence where simpler assumptions fail.

However, copula VAEs remain unexplored in the field of SHM.
Indeed, this is the first work embedding a copula into a VAE for inferring the damage condition of an FOWT system.
We design a Gaussian copula VAE architecture with truncated Gaussian marginals, thus ensuring that Copula samples lie within the desired latent space domain. This truncation requires incorporating rejection sampling within the VAE architecture. 
We implement the proposed methodology for the two-dimensional case study presented in \cite{Navamuel_wes2025}, which aims to infer the presence of damage in the mooring system of an FOWT from motion measurements of its platform.  
We compare the performance of the copula with that of diagonal- and full-covariance Gaussian mixture approaches. 
The results obtained suggest that the copula approach has superior scalability properties in higher dimensions. 


This paper is organized as follows: Section 2 describes the physical problem to solve, including the FOWT dynamics and the data generation process; Section 3 formulates the VAE architecture and associated loss function, describing the different posterior distributions to be explored (i.e., the diagonal and full covariance Gaussian mixtures, and the Gaussian copula), including the sampling strategy. Section 4 presents the specifications for building the VAE architecture, including implementation and training details. Section 5 presents the main results in the form of a comparative analysis. Finally, we draw our conclusions and describe opportunities for future work in Section 6.

\section{Problem description} 

\subsection{Governing Dynamics of FOWTs}
\label{sec:Governing_Dynamics}
\begin{figure}[h!]
    \centering
    \includegraphics[width=0.75\linewidth]{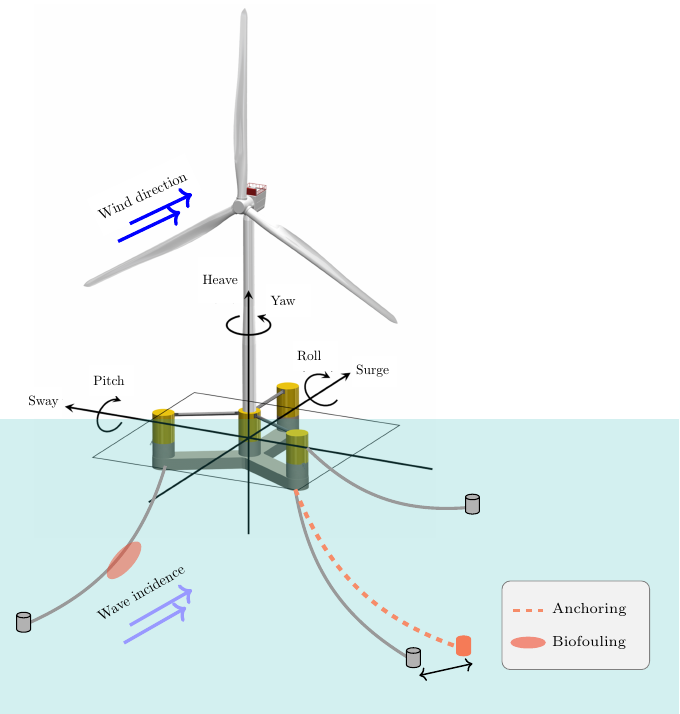}
    \caption{\textit{Schematic representation of the system under study. The figure illustrates the two considered damage types: anchoring (displacement of the line anchor) and biofouling (added mass due to mollusks, algae, or other species adhering to the mooring line). Wind and wave incidence follow the surge direction, which is perpendicular to the plane of the blades.}}
    \label{fig:FOWT_system_with_damage}
\end{figure}

In this paper, we consider an FOWT with a mooring system that anchors the platform to the seabed. In Figure~\ref{fig:FOWT_system_with_damage}, we show a schematic of the system for a particular mooring system formed by three lines that anchor the platform to the system. 
We identify in Figure~\ref{fig:FOWT_system_with_damage} the six degrees of freedom (DOFs) that describe the rigid body motion of the floating platform, namely: sway, surge, heave, pitch, roll, and yaw. The first three DOFs refer to displacements, and the last three correspond to rotations. 

We express the dynamics governing the platform's behavior using a system of equations based on Newton's second law.
A widely employed approach to describe the time domain behavior of FOWTs can be expressed as a rigid body \cite{faltinsen1993sea, jonkman2011dynamics}:
\begin{equation}
    (M_{S}+A_\infty)\Ddot{u}(t)+K_{S}u(t) = \sum F_{\text{ext}}(t, \omega)
    \label{eq:newton_simple}
\end{equation}
Here, $u(t)$ contains the six rigid-body DOFs: surge, sway, heave, roll, pitch, and yaw, and $\Ddot{u}(t)$ is the corresponding vector of accelerations. 
Matrices $M_{S}$ and $K_{S} \in \mathbb{R}^{n_{\text{dof}}\times {n_{\text{dof}}}}$ encode the system's mass and stiffness of the FOWT system, respectively, for the six DOFs ($n_{\text{dof}} = 6$). 
We use the subscript $S$ to denote the system properties.
The matrix $A_\infty \in \mathbb{R}^{n_{\text{dof}} \times n_{\text{dof}}}$  describes the inertial effect of water being accelerated as a consequence of the platform's unsteady motion. This phenomenon is introduced in the model as an added mass at infinite frequency. 
The right-hand term comprises all the external forces that affect the system, which depend upon time $t$ and the wave frequency $\omega$ (in rad/s).

We can extend the expression for the FOWT motion by using Cummins equation \cite{NEWMAN1979221, Hatecke01062015}, which accounts for the fluid memory effects associated with the hydrodynamic radiation:
\begin{equation}
\begin{split}
    (M_S+A_\infty)\Ddot{u}(t) + \int_0^t R_S(t-\tau)\Dot{u}(\tau)d\tau + K_S u(t) = \\ = F_{\text{wave}}(t, \mathbf{w}) + F_{\text{wind}}(t, \mathbf{w}) + F_{\text{visc}}(t, u, \Dot{u}) + F_{\text{moor}}(t, u, \mathbf{w}, S_{\text{moor}}),
\end{split}
    \label{eq:cummins_eq}
\end{equation}
where $F_{\text{wave}}(t, \mathbf{w})$ denotes the wave-induced forces, $F_{\text{wind}}(t)$ denotes the aerodynamic forces, $F_{\text{visc}}$ denotes the viscous hydrodynamic forces and moments owing to fluid viscosity, $F_{\text{moor}}(t, u, \mathbf{w}, S_{\text{moor}})$ denotes the mooring system forces, and $\mathbf{w}$ denotes the environmental loading conditions. 
The retardation functions $R_S(t)$ are related to the frequency-dependent radiation damping coefficients $B(\omega)$ via the Fourier transform:
\begin{equation}
    R_S(t) = \frac{2}{\pi} \int_0^\infty B(\omega) \cos(\omega t) d\omega.
    \label{eq:retardation_function_merged}
\end{equation}
The dependency between the frequency-dependent added mass $A(\omega)$,$A_\infty$, and $R_S(t)$ is described by the following relation:
\begin{equation}
    A(\omega) = A_\infty + \frac{1}{\omega} \int_0^\infty R_S(t) \sin(\omega t) dt.
    \label{eq:added_mass_freq_relation_merged}
\end{equation}
The hydrodynamic coefficients $A(\omega)$ and $B(\omega)$ are typically pre-calculated using boundary element method (BEM) codes (e.g., WAMIT, AQWA). Equation \eqref{eq:cummins_eq} constitutes the forward operator $\mathcal{F}$ to determine the response of the FOWT given the external excitation and the system's inner properties. Simulation tools, such as OpenFAST \cite{jonkman2007dynamics}, solve this equation to calculate the system's response.

\subsection{Representative damage-sensitive features}
Changes in mooring line integrity directly influence the terms representing mooring force $F_{\text{moor}}(t, u, S_{\text{moor}})$ in eq.~\eqref{eq:cummins_eq}. 
This alteration of the mooring system properties $S_{\text{moor}}$ affects the system's response, from which, in practice, only some DOFs are typically measured.
OpenFAST simulations produce 15-minute time-domain signals that represent a certain scenario. For data-storage efficiency, we reduce these data by extracting some representative features that describe the system's behavior.
The obtained key quantities are: $i)$ the mean ($\bar{x}$), $ii)$ the standard deviation (SD), $iii), iv)$ the two main frequencies $f_1$ and $f_{2}$ (Hz) --- one corresponding to the platform and the other to the excitation ---, and  $v)$ the zero$-$th momentum ($m_{0}$).
All of these are directly affected by mooring damage (cf.~\cite{Navamuel_wes2025}) and will thus be used to feed the VAE architecture for mooring damage identification. 
These quantities are defined as follows:

\textbf{$(i)$ Mean: }We define the mean displacement $\bar{x}$ to be:
\begin{equation}
    \bar{x} = \int_{t_0}^{t_f}{x}dt \approx \frac{1}{N}\sum_{i=1}^N{x_i},
\end{equation}
where $N$ indicates the total number of data points. 
A significant fault in a mooring line typically reduces the system’s overall restoring capability in the horizontal plane (surge, sway) and yaw \cite{xu2021design}. 
This results in a noticeable shift in the platform’s mean offset $\bar{x}$ in the affected DOFs.

\textbf{$(ii)$ Standard deviation}: We obtain the standard deviation of the response as:
\begin{equation}
    \text{SD}(x) = \sqrt{\frac{1}{N-1}\sum_{i=1}^N{(x_i-\Bar{x})^2}},
\end{equation}
where we assume the time domain response to be stationary by neglecting its transient state.
Damage often produces a reduction in the stiffness of the mooring system. A “softer” mooring will generally exhibit larger dynamic variations in response to the same excitation, as the platform's ability to resist dynamic loads is compromised.

\textbf{$(iii, iv)$ Platform and loading frequencies:} In the frequency domain, we identify two dominant peak frequencies that account for the platform's main natural frequency and the influence of external loading conditions. 
To obtain them, we calculate the Power Spectral Density (PSD) of the signals~\cite{THEODORIDIS202019}: 
\begin{equation}
    S_x(f) = \lim_{T\to\infty}{\frac{\mathbb{E}\left[\left|F_{x}(f)\right|^2\right]}{2T}},
\end{equation}
where $F_x(f)$ is the Fourier Transform of the time-domain signal for any DOF $x$, as a function of frequency $f$ (Hz). 
We obtain the two dominant peak frequencies as:
\begin{equation}
    f_1 = \argmax_{f\in\left[0, f_{\text{lim}}\right]}S_x(f),
\end{equation}
\begin{equation}
    f_2 = \argmax_{f\in\left[f_{\text{lim}},\infty\right]}S_x(f),
\end{equation}
where we split the natural excitation frequency of each DOF according to a threshold value $f_{\text{lim}}$ to ensure identifying both peaks. A stiffness reduction induced by damage will characteristically lower these natural frequencies, shifting the corresponding low-frequency peak $f_{1}$.
Higher-frequency peaks ($f_2$) might also be affected by changes in mooring conditions due to coupling effects, though they are primarily governed by hydrostatic stiffness ($K_S$) and mass and inertia properties ($M_S, A_\infty$). If $f_2$ represents a dominant wave excitation frequency, it would remain unchanged unless the system's transfer function is altered such that a different excitation frequency becomes dominant in the response.

\textbf{$(v)$ Zero-th momentum: }Finally, to assess the magnitude of the peaks and the intensity of all the frequencies in the spectra, we also measure the zero-th momentum of the PSD as~(\cite{sundar2017ocean}):
\begin{equation}
    m_0=2\pi\int_0^{f_N}{f{S_x}(f)df},
\end{equation}
where $f_N$ is the Nyquist frequency in Hz~\cite{robinson1991sampling}.
As this feature represents the power-weighted average frequency (first spectral moment), it is sensitive to shifts in the overall energy distribution. If mooring damage leads to an increase in low-frequency response energy (e.g., due to the reduction in natural frequency $f_1$), the value of this $m_0$ is expected to decrease. Conversely, if damage excites relatively more high-frequency content or significantly broadens the spectrum, the effect on $m_0$ would depend on the specifics of the spectral shape change. 


\subsection{Synthetic data generation}
\label{Sec:Data_generation}
This section describes the methodology employed for creating a labeled synthetic dataset needed to train the VAE. 

We denote $\mathcal{F}$ the (exact) solution operator of the forward problem (see eq.\eqref{eq:cummins_eq}), i.e., given the system's condition and the affecting environmental conditions, $\mathcal{F}$ yields its motion response described by six DOFs. 
In practice, this operator is unknown, and an approximation is needed to generate labeled data to train NNs. 
Here, we employ the NREL's open-source wind turbine simulation tool OpenFAST~\cite{openfast}, which accounts for the influence of aerodynamic~\cite{jonkman2015aerodyn} and hydrodynamic~\cite{jonkman2014hydrodyn} excitations on the response of the floating platform. 
OpenFAST is a highly accurate and reliable tool to perform numerical simulations of FOWTs~\cite{yang2021investigation, reig2024efficient, rinker2020comparison}.
From now on, we will refer to the OpenFAST solver as $\mathcal{F}$ since it represents a sufficiently accurate approximation to the exact operator. 

We simulate a 10-MW floating offshore wind turbine (FOWT) under a wide range of environmental conditions.
We sample each condition uniformly at random in the following ranges:  significant wave height ($H_{S}\in [2,15]$ m), peak period ($T_{P} \in [1,15]$ s), and wind velocity ($W_{V}\in [1,30]$ m/s).
For each condition, we introduce a mooring line fault by considering two damage mechanisms: biofouling and anchor point slippage. 
For simplicity in the analysis, we consider both damage types affecting the same mooring line. 
We introduce biofouling as a simultaneous percentage increase in the line's mass per unit length and its diameter, ranging from a $0\%$ (undamaged) to a $10\%$ mass increase (maximum damage level). 
To simulate anchoring, we modify the coordinates of the anchorage point to represent displacement in the wave direction, ranging from $0$ meters (undamaged) to a 20-meter shift of the line's anchor (maximum damage level), assuming a flat seabed. 
To simulate the damage scenarios, we modify the corresponding parameters in OpenFAST. We define a severity coefficient for each damage type in the form of a bi-dimensional vector $\mathbf{z}= \{z_{1}, z_{2}\} \in [0,1]^{2}$, where $0$ indicates undamaged and $1$ refers to the maximum damage level maximum damage level (corresponding to $10\%$ mass increase for biofueling and 20m line anchor shift). Here, $z_{1}$ refers to biofouling and $z_{2}$ to anchoring. 
We randomly sample the severity coefficients from a folded Gaussian distribution, ensuring that the dataset is primarily composed of healthy and low-severity scenarios that reflect realistic situations for early damage identification.

\begin{figure}[h!]
    \centering
    \begin{subfigure}[b]{0.48\textwidth}
        \centering
        \includegraphics[width=\textwidth]{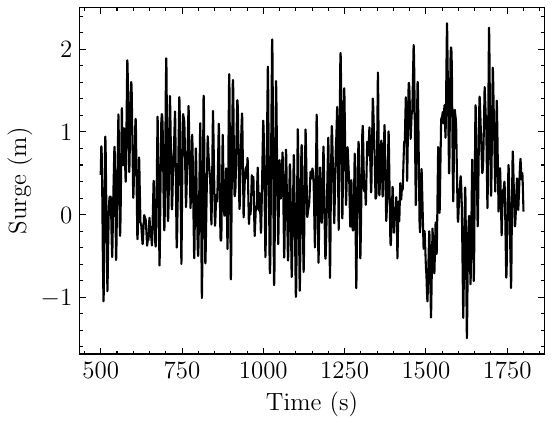}
        \caption{}
        \label{fig:surge_data_time}
    \end{subfigure}
    \hfill 
    \begin{subfigure}[b]{0.48\textwidth}
        \centering
        \includegraphics[width=\textwidth]{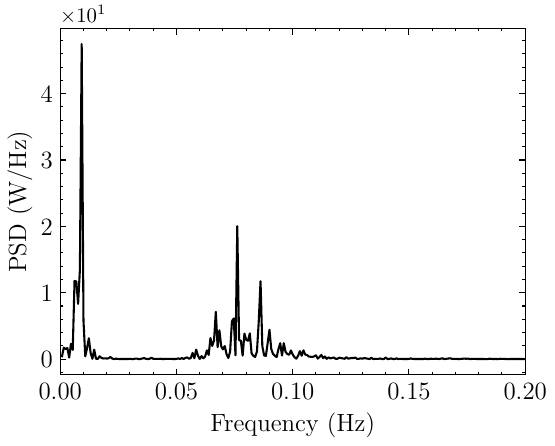}
        \caption{}
        \label{fig:surge_data_freq}
    \end{subfigure}
    \caption{\textit{ Example of the generated response data of the surge DOF for one scenario in the (a) time and (b) frequency domains.}}
    \label{fig:Surge_data_plot}
\end{figure}

We run a total of $60,000$ time-domain simulations, each capturing 30 minutes of FOWT dynamics, from which we extract the five relevant motion response features (see Section \ref{sec:Governing_Dynamics}). 
For each scenario, we encode the selected features in vector $\mathbf{m}\in \mathbb{R}^{n_{m}}$, containing the five features for each of the $n_{\text{dofs}}$ measured DOFs ($n_{m} = 5\times n_{\text{dofs}}$). 
In practice, $n_{\text{dofs}}$ depends on the available sensors.
Analogously, the affecting environmental conditions are encoded as $\mathbf{w} = \{H_{S}, T_{P}, W_{V}\}$.
For any pair $\{\mathbf{m}, \mathbf{w}\}$, vector $\mathbf{z}$ denotes the true damage condition to be estimated with the VAE. 
In Figure \ref{fig:Surge_data_plot}, we show the response in the surge DOF for one synthetic scenario. 
\section{Inverse problem for the FOWT mooring system assessment}
\label{Sec:inverse_problem_formulation}
This section describes the methodology we propose for solving the damage identification inverse problem. We first introduce the different components of the VAE architecture (i.e., the encoder, the decoder, and the sampling layer). We formulate a function that accounts for both data misfit and measurement uncertainty, and we propose a strategy for loss function approximation and optimization. 

\subsection{VAE architecture}
\label{Sec:VAE_arch}
When damage occurs, the FOWT mooring system's motion response is affected.
Identifying the mooring system's state based on measurements of its motion response $\mathbf{m} \in \mathcal{M}$  (and environmental excitation $\mathbf{w} \in \mathcal{W}$ ) is an inverse problem. 
In what follows, we denote with $\mathcal{I}$ the (inverse) operator which maps observation data onto a set $\mathcal{Z}$ of damage descriptors or features $\mathbf{z}\in\mathcal{Z}$, i.e., $\mathcal{I}: \mathcal{M} \times \mathcal{W}\rightarrow \mathcal{Z}$. Here, $\mathbf{z}$ contains $D \geq 2$ features.

To describe the uncertainty in the measured motion responses $\mathbf{m}$, environmental conditions (wind and wave excitation) $\mathbf{w}$, and unknown system damage condition $\mathbf{z}$, we model them as random variables.
These quantities are related through the equation:
\begin{equation}
\mathbf{m} = \mathcal{F}(\mathbf{z}, \mathbf{w}) + \bm{\epsilon},
\label{eq:noise_forward}
\end{equation}
where $\bm{\epsilon}$ is an additive noise term that represents data acquisition and modeling errors, and $\mathcal{F}$ denotes the exact forward operator that describes the motion response $\mathbf{m}$ of the FOWT system.

\begin{figure*}[ht]
		\centering
		\includegraphics[width=0.9\linewidth]{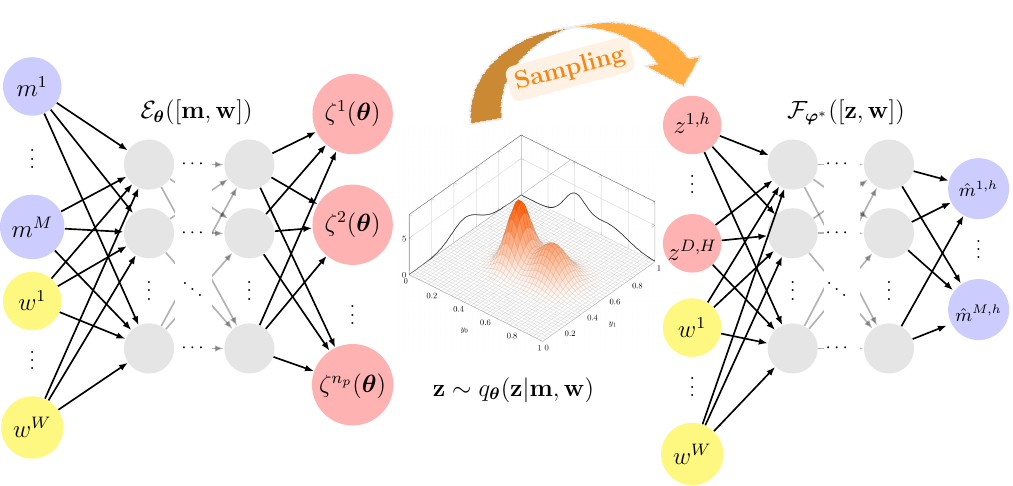}
		\caption{\textit{Variational autoencoder architecture. The encoder estimates the properties that describe the posterior PDF of the latent space (damage condition features), $\mathbf{z} \sim q_{\bm{\theta}}(\mathbf{z}|\mathbf{m}, \mathbf{w})$. A sampling layer draws $H$ random samples from the distribution, which are then fed to the optimal forward operator $\mathcal{F}_{\bm{\varphi}^{*}}$. The output layer yields the reconstruction of the input measurements, $\hat{\mathbf{m}}_{i}^{h} = \mathcal{F}_{\bm{\varphi}^{*}} \circ \mathcal{E}_{\bm{\theta}}(\mathbf{m}_{i}^{h}, \mathbf{w}_{i})$, for each sample $h = 1,..., H$. }}
		\label{fig:GMM_InverseForwardArch}
	\end{figure*}

Following~\cite{Navamuel_wes2025}, we design a Bayesian VAE architecture to estimate the posterior distribution of the estimated damage properties $\mathbf{z}$ given measurements of the system's response $\mathbf{m}$ and of the environmental conditions $\mathbf{w}$. 
This architecture is formed by the following components (see Figure~\ref{fig:GMM_InverseForwardArch} for a schematic): an encoder, a sampling layer, and a decoder.  
The encoder is a fully-connected neural network (NN) $\mathcal{E}_{\bm{\theta}}$, parametrized by $\bm{\theta}$, which maps the motion $\mathbf{m}$ and environmental condition $\mathbf{w}$ measurements to the parameters $\bm{\zeta}_{\bm{\theta}}\in \mathbb{R}^{n_{p}}$ of a parametric posterior probability density function (PDF) denoted as $q_{\bm{\zeta}_{\bm{\theta}}}(\mathbf{z}|\mathbf{m},\mathbf{w})$. 
Symbol $q(\cdot)$ represents a computationally tractable approximation of a true PDF, $p(\cdot)$. Particularly, $q_{\bm{\zeta}_{\bm{\theta}}}(\mathbf{z}|\mathbf{m},\mathbf{w})$ approximates the true and unknown posterior PDF $p(\mathbf{z}|\mathbf{m},\mathbf{w})$.
For simplicity of notation, we will write $q_{\bm{\theta}}(\cdot)$ in place of $q_{\bm{\zeta}_{\bm{\theta}}}(\cdot)$ throughout the paper. The density $q_{\bm{\theta}}(\mathbf{z}|\mathbf{m},\mathbf{w})$ is taken as an approximation of the true unknown posterior distribution $p(\mathbf{z}|\mathbf{m}, \mathbf{w})$.

The sampling layer draws random samples from the approximate distribution $q_{\bm{\theta}}$. 
The decoder $\mathcal{F}_{\bm{\varphi}^{*}}$ is a pretrained fully-connected NN parametrized by $\bm{\varphi}^{*}$ that reconstructs the system's motion response. 
It acts as a surrogate that substitutes the exact forward operator $\mathcal{F}$ (or its computationally expensive OpenFAST solver approximation), enabling the timely execution of simulations required to learn the optimal encoder.  
Details on the training step of the decoder NN can be found in~\cite{Navamuel_wes2025}.

\subsection{ELBO calculation}
To train the encoder NN (i.e., $\mathcal{E}_{\bm{\theta}}$), we freeze the optimal decoder parameters ($\bm{\varphi}^{*}$) and learn the optimal parameter set $\bm{\theta}^{*} $.
After completing the training, for any observed input pair $\{ \mathbf{m}, \mathbf{w}\}$, we obtain the set of properties $\bm{\zeta}_{\bm{\theta}^{*}} = \mathcal{E}_{\bm{\theta}^{*}}([\mathbf{m},\mathbf{w}])$ parameterizing the approximate posterior PDF of the damage condition properties, $\mathbf{z} \sim q_{\bm{\theta}^{*}}(\mathbf{z}|\mathbf{m},\mathbf{w})$. 
We assume that the unknown measurement error $\bm{\epsilon}$ is distributed according to a known PDF $p_{\epsilon}(\cdot)$ and that it is independent from the unknown damage condition $\mathbf{z}$ \cite{BayesInverse_Goh2021}. 
Thus, according to eq.~\eqref{eq:noise_forward}, we can express the likelihood $p(\mathbf{m}|\mathbf{w}, \mathbf{z})$ as: 
\begin{equation}
p(\mathbf{m}|\mathbf{w}, \mathbf{z}) = p_{\epsilon} \left(\mathbf{m} - \mathcal{F}_{\bm{\varphi}^{*}}([\mathbf{z},\mathbf{w}])\right).
\label{eq:p_lklhd}
\end{equation}


Although $\mathbf{z}$ is unknown, its uncertainty can be modeled using a conditional probability distribution $p(\mathbf{z}|\mathbf{m}, \mathbf{w})$. 
For any damage condition $\mathbf{z}$, the conditional PDF $p(\mathbf{z}|\mathbf{m},\mathbf{w})$ is the target posterior distribution that solves the inverse problem under the Bayesian framework. 
Bayes' theorem \cite{joyce2003bayes} implies that: 
\begin{equation}
p(\mathbf{z}| \mathbf{m}, \mathbf{w}) \propto p(\mathbf{m}|\mathbf{w}, \mathbf{z}) \cdot p(\mathbf{z}),
\label{eq:post_bayes}
\end{equation}
where $p(\mathbf{z})$ represents the prior PDF of the unknown damage condition $\mathbf{z}$, and $p(\mathbf{m}|\mathbf{z}, \mathbf{w})$ is the likelihood model that expresses the interrelation between the observations and the damage condition. 

Due to the intractability of the true posterior, we employ the encoder operator $\mathcal{E}_{\bm{\theta}}$ to estimate the parameters of an approximate PDF denoted as $q_{\bm{\theta}}(\mathbf{z}|\mathbf{m}, \mathbf{w})$, given the motion $\mathbf{m}$ and environmental excitation $\mathbf{w}$ measurements.
Realizations from the approximate posterior $q_{\bm{\theta}}(\mathbf{z}|\mathbf{m}, \mathbf{w})$ represent samples of the damage condition $\mathbf{z}$ that likely produced by the unknown true posterior $p(\mathbf{z}|\mathbf{m}, \mathbf{w})$. 
We remark that all the realizations $\{\mathbf{z}_{h}\}_{h =1}^{H}$ share the same operating conditions $\mathbf{w}$ (recall Section~\ref{Sec:Data_generation}). 

Our objective is to maximize the likelihood of the data while at the same time minimizing the discrepancy between the true posterior $p(\mathbf{z}|\mathbf{m},\mathbf{w})$ and the approximate posterior $q_{\bm{\theta}}(\mathbf{z}|\mathbf{m},\mathbf{w})$ obtained from $\mathcal{E}_{\bm{\theta}}$.
To achieve this goal, we select the evidence lower bound (ELBO) as the loss function, a standard choice in VAE algorithms~\cite{VInference_ELBO2017}. 
For any distributional model $q_{\bm{\theta}}(\mathbf{z}|\mathbf{m},\mathbf{w})$, the evidence, $\log{p}(\cdot)$, is given by~\cite{VAE_tutorial}:
\begin{equation}
\begin{split}
\log p(\mathbf{m},\mathbf{w}) = \mathbb{E}_{q_{\bm{\theta}}}[\log p(\mathbf{m},\mathbf{w})] &=
\mathbb{E}_{q_{\bm{\theta}}}\left[ \log \left[ \frac{p(\mathbf{m},\mathbf{w}|\mathbf{z})}{p(\mathbf{z}|\mathbf{m},\mathbf{w})}\right] \right] \\
&= \mathbb{E}_{q_{\bm{\theta}}}\left[\log \left[ \frac{p(\mathbf{m},\mathbf{w}|\mathbf{z})q_{\bm{\theta}}(\mathbf{z}|\mathbf{m},\mathbf{w})}{q_{\bm{\theta}}(\mathbf{z}|\mathbf{m},\mathbf{w})p(\mathbf{z}|\mathbf{m},\mathbf{w})} \right] \right] \\
&= \underbrace{\mathbb{E}_{q_{\bm{\theta}}}\left[\log \left[ \frac{p(\mathbf{m},\mathbf{w}|\mathbf{z})}{q_{\bm{\theta}}(\mathbf{z}|\mathbf{m},\mathbf{w})} \right] \right]}_{\textcolor{blue}{\text{ELBO}}} + \underbrace{\mathbb{E}_{q_{\bm{\theta}}}\left[\log \left[ \frac{q_{\bm{\theta}}(\mathbf{z}|\mathbf{m},\mathbf{w})}{p(\mathbf{z}|\mathbf{m},\mathbf{w})} \right] \right]}_{\textcolor{blue}{D_{\text{KL}}(q_{\bm{\theta}}(\mathbf{z}| \mathbf{m},\mathbf{w})||p(\mathbf{z}|\mathbf{m},\mathbf{w}))}},
\label{eq:ELBO_derivation}
\end{split}
\end{equation}
where $D_{\text{KL}}$ is the Kullback-Leibler divergence between two PDFs $p$ and $q$, and is defined as~\cite{VAE_tutorial}: 
\begin{equation}
D_{\text{KL}}[p(x)||q(x)] = \int p(x) \ \text{log} \frac{p(x)}{q(x)}dx.
\end{equation}
Note that $D_{\text{KL}}[p||q]$ is zero if and only if $p = q$~\cite{KL_Divergence}.  
Since $D_{\text{KL}}$ is non-negative by definition, the first term in eq.~\eqref{eq:ELBO_derivation} constitutes a lower bound. 
Rearranging terms, we define the ELBO loss as: 
\begin{equation}
\begin{split}
\log p(\mathbf{m},\mathbf{w}) - D_{\text{KL}}[q_{\bm{\theta}}(\mathbf{z}|\mathbf{m},\mathbf{w}) || p(\mathbf{z}|\mathbf{m},\mathbf{w})]  \\ &= \mathbb{E}_{q_{\bm{\theta}}}[\log p(\mathbf{m},\mathbf{w}, \mathbf{z}) - \log q_{\bm{\theta}}(\mathbf{z}|\mathbf{m},\mathbf{w})] \\
&= \mathbb{E}_{q_{\bm{\theta}}}[\log (p(\mathbf{m}|\mathbf{w},\mathbf{z})p(\mathbf{w})p(\mathbf{z}))] - \mathbb{E}_{q_{\bm{\theta}}}[\log q_{\bm{\theta}}(\mathbf{z}|\mathbf{m},\mathbf{w})] \\
&= \mathbb{E}_{q_{\bm{\theta}}}[\log p(\mathbf{m},\mathbf{w}|\mathbf{z})] + \mathbb{E}_{q_{\bm{\theta}}}[\log p(\mathbf{z})] - \mathbb{E}_{q_{\bm{\theta}}}[\log q_{\bm{\theta}}(\mathbf{z}|\mathbf{m},\mathbf{w})] \\ &= \text{ELBO}(\bm{\theta}),
\end{split}
\label{eq:ELBO_expectations}
\end{equation}
where we assume that the operating conditions $\mathbf{w}$ and the damage properties $\mathbf{z}$ are independent (i.e., $p(\mathbf{w}, \mathbf{z}) = p(\mathbf{w})\cdot p(\mathbf{z})$). 
Note that we can disregard the factor $ p(\mathbf{w})$ in the third line of eq.~\eqref{eq:ELBO_expectations} as it is independent of $\mathbf{z}$.

\subsection{ELBO approximation through sampling}
\label{Sec:ELBO_loss}
For a certain observation $\{\mathbf{m},\mathbf{w}\}$, we draw $H$ samples from the posterior and approximate the ELBO function via sample average approximation~\cite{SampleAverageApprox2014guide}:
\begin{equation}
\text{ELBO}(\bm{\theta}) \approx \frac{1}{H} \sum_{h = 1}^{H} [\underbrace{\log p(\mathbf{m},\mathbf{w}|\mathbf{z}^{h})}_{\textcolor{blue}{\text{Likelihood}}}+ \underbrace{\log p(\mathbf{z}^{h})}_{\textcolor{blue}{\text{Prior}}} - \underbrace{\log q_{\bm{\theta}}(\mathbf{z}^{h}|\mathbf{m},\mathbf{w})}_{\textcolor{blue}{\text{Approx. posterior}}}].
\label{eq:ELBO_loss1}
\end{equation} 
The first term in eq. \eqref{eq:ELBO_loss1} accounts for the data misfit, which is the error between the true measurements and the reconstructions provided by $\mathcal{F}_{\bm{\varphi}^{*}}$. 
The second term refers to the prior, which we assume to follow a bounded uniform distribution $p(\mathbf{z}) \sim \mathcal{U}[b_{\text{low}},b_{\text{up}}]^{D}$ with lower and upper bounds $b_{\text{low}}$ and $b_{\text{up}}$, respectively. The last term measures the probability that the $h$-th sample belongs to the estimated distribution $q_{\bm{\theta}}(\mathbf{z}|\mathbf{m},\mathbf{w})$. 

In this work, we assume the noise follows a Gaussian distribution, $p(\bm{\epsilon}) = \mathcal{N}(0, \Gamma)$, where $\Gamma = \text{diag}(\beta^{2})$ is a diagonal matrix containing the noise variances, and the parameter $\beta$ represents the noise level.
We can thus rewrite eq.~\eqref{eq:p_lklhd} as: 
\begin{equation}
p(\mathbf{m}|\mathbf{w}, \mathbf{z})= \frac{1}{(2\pi)^{M/2}|\Gamma|^{1/2}} \exp\left(-\frac{1}{2}(\mathbf{m}  - \mathcal{F}_{\bm{\varphi}^{*}}([\mathbf{z},\mathbf{w}]))^{\top} \Gamma^{-1} ( \mathbf{m} - \mathcal{F}_{\bm{\varphi}^{*}}([\mathbf{z},\mathbf{w}]))\right), 
\label{eq:conditional_gaussianoise}
\end{equation}
By substituting the likelihood from eq.~\eqref{eq:conditional_gaussianoise} into eq.~\eqref{eq:ELBO_loss1}, we finally express the ELBO  as: 
\begin{equation}
\text{ELBO}(\bm{\theta}) \approx \frac{1}{H} \sum_{h = 1}^{H} \left[-\frac{1}{2} (\mathbf{m}  -\mathcal{F}_{\bm{\varphi}^{*}}([ \mathbf{z}^{h},\mathbf{w}]))^{\top} \Gamma^{-1} ( \mathbf{m} - \mathcal{F}_{\bm{\varphi}^{*}}([\mathbf{z}^{h},\mathbf{w}])) + \log p(\mathbf{z}^{h}) - \log q_{\bm{\theta}}(\mathbf{z}^{h}|\mathbf{m},\mathbf{w})\right].
\label{eq:ELBO_loss}
\end{equation} 
The second term in eq.~\eqref{eq:ELBO_loss} can be neglected by directly constraining the Gaussian mixture density function to the desired interval. 


The next subsection explores different configurations for the approximate posterior $q_{\bm{\theta}}(\mathbf{z}|\mathbf{m}, \mathbf{w})$ in eq.~\eqref{eq:ELBO_minimization}, including the related sampling strategies (see Figure \ref{fig:GMM_InverseForwardArch}).

\subsection{Posterior distribution approximation}
\label{sec:Posteriors}
A key feature of the VAE architecture is the approximate posterior distribution $q_{\bm{\theta}}(\mathbf{z}|\mathbf{m}, \mathbf{w})$. 
It must be sufficiently expressive to capture complex probability maps in the latent space while being computationally efficient. 
We explore two different approaches for approximating the unknown true posterior distribution $p(\mathbf{z}|\mathbf{m},\mathbf{w})$: a Gaussian mixture and a Copula model. 
In this section, we define both the approximate PDF $q_{\bm{\theta}}(\mathbf{z}|\mathbf{m},\mathbf{w})$ to be embedded in eq.~\eqref{eq:ELBO_minimization}, and the corresponding sampling strategy required to feed the decoder (see Figure \ref{fig:GMM_InverseForwardArch}).

\subsubsection{Gaussian mixture}
\label{sec:Gaussian_Mixture_posterior}
Let  $q_{\bm{\theta}}(\mathbf{z}|\mathbf{m}, \mathbf{w})$ be a Gaussian mixture (GM)~\cite{yu2008multimode} with $K$ components, namely: 
\begin{equation}
    q^{\text{GM}}_{\bm{\theta}}(\mathbf{z}|\mathbf{m}, \mathbf{w}) := \sum_{k=1}^{K} \alpha_{k} \mathcal{G}(\mathbf{z} |\bm{\zeta}_k(\bm{\theta})),
\label{eq:GM_expression}
\end{equation}
where $\alpha_{k}$ is the weight for the $k$-th Gaussian component, and $\mathcal{G}\left(\mathbf{z} |\bm{\zeta}_k(\bm{\theta})\right)$ is the corresponding multivariate Gaussian PDF with mean vector $\bm{\mu}_{k}$ and covariance matrix $\Sigma_{k}$, i.e.:
\begin{equation}
    \mathcal{G}(\mathbf{z} | \bm{\mu}_{k}, \Sigma_{k}) = \frac{1}{(2\pi)^{D/2}|\Sigma_{k}|^{1/2}}\exp \left(- \frac{1}{2}(\mathbf{z}-\bm{\mu}_{k})^{\top}\Sigma_{k}^{-1}(\mathbf{z}-\bm{\mu}_{k}) \right).
\label{eq:Gaussian_PDF}
\end{equation}

We assume full covariance matrices for the Gaussian components. 
For each component, we parameterize the covariances using the Cholesky factorization $\bm{\Sigma}^{\text{full}}_{k} = \mathbf{L}_{k}\mathbf{L}_{k}^{\top}$, where $\mathbf{L}_{k}$ is a lower triangular matrix. 
Therefor, for a $D$-dimensional space, we must estimate a total of $n_{L} = \frac{D(D-1)}{2}$ entries to fill $\mathbf{L}_{k}$.
This formulation is more general than the one from~\cite{Navamuel_wes2025, Rodriguez2023}, which considers diagonal covariance structures and thus cannot account for correlations among latent variables. 
At the same time, it is also more expensive due to the cost of dense linear algebra operations and an increased number of distribution parameters
Hence, the inverse operator $\mathcal{E}_{\theta}(\cdot)$ in Figure~\ref{fig:GMM_InverseForwardArch} must estimate a tensor $\bm{\mu} \in \mathbb{R}^{K\times D}$ with the $D$-dimensional means of the $K$ components, a tensor $\mathbf{l} \in \mathbb{R}^{K\times n_{L}}$ with the non-zero lower triangular matrix entries for all the components, and a vector $\bm{\alpha}$ with $K-1$ weights. 
Thus, the parameter vector $\bm{\zeta}(\bm{\theta})\in \mathbb{R}^{n_{p}}$ to be estimated contains the flattened tensors: $\bm{\zeta}(\bm{\theta}) = \{\bm{\mu}(\bm{\theta}), \mathbf{l}(\bm{\theta}), \bm{\alpha}(\bm{\theta})\}$. 
For simplicity of notation, from now on we omit the dependence of the distributional parameters $\bm{\zeta}(\bm{\theta})$ on the NN parameters $\bm{\theta}$. 

The damage condition features $\mathbf{z}$ are bounded in a fixed domain $\mathcal{A} = [b_{\text{low}}, b_{\text{up}}]$, where  $b_{\text{low}}$ and $b_{\text{up}}$ stand for the lower and upper domain bounds, respectively.
We recall that here $\mathcal{A}$ corresponds to the D-dimensional unit hypercube, i.e., $\mathcal{A} = [0,1]^D$ (cf.~Section~\ref{Sec:Data_generation}).
However, the Gaussian mixture PDF is defined over all $\mathbb{R}^{D}$ and thus has non-zero support outside the feasibility domain $\mathcal{A}$. 
To address this issue, we truncate the PDF of the Gaussian mixture $q^{\text{GM}}_{\bm{\theta}}$ by restricting its support as follows: We multiply $q^{\text{GM}}_{\bm{\theta}}$ by the indicator function of $\mathcal{A}$, i.e., we set $\tilde{q}^{\text{GM}}_{\bm{\theta}}(\mathbf{z} | \mathbf{m}, \mathbf{w}) = q^{\text{GM}}_{\bm{\theta}}(\mathbf{z}|\mathbf{m},\mathbf{w}) \bm{1}_{\mathcal{A}}(\mathbf{z})$. This ensures that all samples from this density belong to $\mathcal{A}$ with probability $1$. Such $\tilde{q}^{\text{GM}}_{\bm{\theta}}$ is, however, unnormalised (it is not a PDF since it does not integrate to $1$). The normalising constant $C$ is given by:
\begin{equation}
    C(\bm{\theta}, \mathbf{m}, \mathbf{w})  = \int_{\mathcal{A}} q_{\bm{\theta}}(\mathbf{z}|\mathbf{m}, \mathbf{w})d\mathbf{z} = \int_{b_{\text{low}}}^{b_{\text{up}}} q_{\bm{\theta}}(\mathbf{z}|\mathbf{m}, \mathbf{w})dz_{1}...dz_{D}
    \label{eq:Normalizing_constant}
\end{equation}
Since the above integral does not admit a closed-form expression, we use a plug-in Monte Carlo (MC) estimator to approximate $C(\bm{\theta}, \mathbf{m}, \mathbf{w})$ in the ELBO loss (cf. eqn~\eqref{eq:ELBO_loss}). This MC estimator reads:
\begin{equation}
    \int_{\mathcal{A}} q_{\bm{\theta}}(\mathbf{z}|\mathbf{m}, \mathbf{w})d\mathbf{z} \approx  \frac{\text{vol}(\mathcal{A})}{N_{\text{MC}}}\sum_{j=1}^{N_{\text{MC}}}q_{\bm{\theta}}(\mathbf{u}_{j}|\mathbf{m},\mathbf{w}) \approx C(\bm{\theta}, \mathbf{m}, \mathbf{w}),
\end{equation}
where $\text{vol}(\mathcal{A})$ corresponds to the volume of $\mathcal{A}$, and $\mathbf{u}_{j} \sim \mathcal{U}(\mathcal{A})$ represents samples drawn uniformly from $\mathcal{A}$. 

In this work, we truncate the Gaussian mixture PDF by applying rejection sampling~\cite{bauer2019resampled}. We chose this strategy to ensure compatibility with the TensorFlow software.
At each training iteration, we thus keep resampling until we obtain H samples belonging to $\mathcal{A}$, i.e., $\mathbf{z}^{h}\in \mathcal{A}, \ h = 1,..., H$. 
Algorithm~\ref{alg:rejection_sampling} describes the rejection sampling step. 
\begin{algorithm}[h!]
\caption{\textit{Truncated Gaussian mixture samples via rejection sampling}}
\label{alg:rejection_sampling}
\begin{algorithmic}[1]
\Require Unconstrained GM $q^{\text{GM}}_{\bm{\theta}}(\mathbf{z}|\mathbf{m},\mathbf{w})$, domain $\mathcal{A}$ with volume $\text{vol}(\mathcal{A})$, number of samples $H$

\State Draw initial batch $\mathcal{S} = \{\mathbf{z}^1, \dots, \mathbf{z}^H\}$ from $q^{\text{GM}}_{\bm{\theta}}(\mathbf{z}|\mathbf{m},\mathbf{w})$
\State Identify the indices corresponding to invalid samples, i.e., those outside of $\mathcal{A}$: $\mathcal{V} = \{j \mid \mathbf{z}^{j} \notin \mathcal{A}\}$
\While{$\mathcal{V} \neq \emptyset$}
    \State Draw a new batch $\mathcal{S}_{\text{new}} = \{\mathbf{z}'^1, \dots, \mathbf{z}'^H\}$ from $q^{\text{GM}}_{\bm{\theta}}(\mathbf{z}|\mathbf{m},\mathbf{w})$
    \For{each index $j \in \mathcal{V}$}
        \State $\mathbf{z}^{j} \leftarrow \mathbf{z}'^{j}$
    \EndFor
    \State Re-identify the indices corresponding to invalid samples, i.e., those outside of $\mathcal{A}$: $\mathcal{V} = \{j \mid \mathbf{z}^j \notin \mathcal{A}\}$
\EndWhile
\State \textbf{Output}: A set of $H$ valid samples $\mathcal{S} \subset \mathcal{A}$
\end{algorithmic}
\end{algorithm}

\subsubsection{Gaussian copula}
\label{sec:Gaussian_Copula_posterior}
In this section, we describe the copula-based posterior approximation. In the context of VAE for FOWT damage identification, this is a novel approach that overcomes the scalability limitations of full-covariance GMs~\cite{tagasovska2019copulas}.

Any multivariate joint distribution can be represented by its marginal distributions and a corresponding copula, which uniquely describes the dependence structure between the variables~\cite{sklar1973random}.
This decomposition is a key feature of copulas since they permit the separate modeling of marginals and correlation dependence. 
Such flexibility is particularly advantageous for constructing complex multivariate models.
Copulas can adopt various forms, mainly divided into (a) parametric (e.g., Gaussian, Archimedean), or (b) non-parametric (e.g., Clayton, Gumbel) \cite{joe1993parametric}. 
For further reading, we refer the reader to the book by Nelsen~\cite{nelsen2006introduction} on copula theory and to the book by Joe~\cite{joe2014dependence} on statistical modelling with copulas.

We can express a joint distribution in terms of the cumulative distribution functions (CDFs) of $D$ uniformly distributed random variables and a copula function $\mathcal{C}: [0,1]^{D}: \rightarrow [0,1]$ according to Sklar's theorem \cite{Copulas_intro1}: 
\begin{equation}
    F_{\mathbf{X}} (x_{1}, ..., x_{D}) := \mathcal{C}(F_{1}(x_{1}), ... F_{D}(x_{D})) = \mathcal{C}(u_{1}, ..., u_{D}),
\end{equation}
where $ \mathbf{F} = \{F_{i}(x_{i})\}_{i=1}^{D}$ contains the marginal CDFs of each variable, which produce uniformly distributed random variables $\mathbf{u} = \{u_{1}, ..., u_{D}\}$. 
The PDF of the copula can be obtained as:
\begin{equation}
    c(\mathbf{u}) := \frac{\partial^{D}\mathcal{C}(u_{1},..., u_{D})}{\partial u_{1}, ..., \partial{u_{D}}}.
\end{equation}
This enables obtaining the joint PDF as: 
\begin{equation}
    f_{\mathbf{X}}(\mathbf{x}) := c(\mathbf{F}(\mathbf{x}))\prod_{i=1}^{D}{f_{d}(x_{d})},
\label{eq:joint_copula_density}
\end{equation}
where $f_{d}(x_{d}), \ d = 1,..., D$ correspond to the marginal PDFs of the random variables $\mathbf{X}$, and $c(\cdot)$ is the copula density function that applies to the CDFs of the input variables. 

In this work, we employ a Gaussian copula as it yields a closed-form expression for its density. Other options are also possible and can be accommodated within our framework with minor adjustments.  
By using a copula, we can express the approximate posterior $q_{\bm{\theta}}(\mathbf{z}|\mathbf{m}, \mathbf{w})$ as~\cite{tran2015copula}: 
\begin{equation}
    q^{\text{Cop}}_{\bm{\theta}}(\mathbf{z}|\mathbf{m}, \mathbf{w}) := c(\mathbf{F}(\mathbf{z}|\mathbf{m}, \mathbf{w}))\prod_{d=1}^{D}{q_{d}(z_{d}|\mathbf{m}, \mathbf{w})},
\label{eq:joint_copula_density}
\end{equation}
where $\mathbf{F}(\mathbf{z}|\mathbf{m}, \mathbf{w})$ is a vector function whose $d$-th entry is the marginal posterior CDF of the $d$-th component of $\mathbf{z}$ and $\{q_{d}(z_d|\mathbf{m},\mathbf{w})\}_{d=1}^{D}$ are the corresponding marginal posterior PDFs. Note that since $F_d$ is the (marginal posterior) CDF of $z_{d}$, it holds that $\mathbf{F}(\mathbf{z} |\mathbf{m}, \mathbf{w})$ is a random vector with uniform marginals.


The zero-mean Gaussian Copula $\mathcal{C}_{\Phi}$ is parametrized by a correlation matrix $\hat{\bm{\Sigma}} \in \mathbb{R}^{D \times D}$, such that: 
\begin{equation}
    \mathcal{C}_{\Phi}(u_{1}, ... u_{D}) = \Phi_{\hat{\bm{\Sigma}}}\left(F_{1}^{-1}(u_{1}), ..., F_{D}^{-1}(u_{D})|\hat{\bm{\Sigma}}\right), 
\end{equation}
where $\Phi_{\hat{\bm{\Sigma}}}(\cdot|\hat{\bm{\Sigma}})$ denotes the $D$-dimensional Gaussian CDF with correlation matrix $\hat{\bm{\Sigma}}$, and $F_{d}^{-1}(\cdot)$ indicates the marginal posterior inverse CDF.  
The Gaussian copula density function $c_{\Phi}$ is given by: 
\begin{equation}
\begin{split}
c_{\Phi}(u_{1}, ..., u_{D}) &= \frac{\partial ^{D} C_{\Phi}(u_{1},..., u_{D})}{\partial u_{1} \cdot \cdot \cdot \partial u_{D}} \\
&= \frac{1}{\sqrt{2\pi |\hat{\bm{\Sigma}}|}}\exp \left( - \frac{1}{2}\left( \mathbf{F}^{-1}(\mathbf{u})\right)^{\top} (\hat{\bm{\Sigma}} ^{-1} - \mathbf{I})\mathbf{F}^{-1}(\mathbf{u}) \right),
\end{split}
\label{eq:copula_term}
\end{equation}
where $\mathbf{F}^{-1}(\mathbf{u}) = \mathbf{F}^{-1}(\mathbf{F}(\mathbf{z}|\mathbf{m},\mathbf{w}))$ is a vector valued function whose entries are the $D$ inverse marginal CDFs, the vector $u = [u_d]_{d=1}^D \in [0,1]^D$, $\hat{\bm{\Sigma}}$ is the copula covariance matrix, and $\mathbf{I}$ is the identity matrix.


We assume truncated Gaussian PDFs for the marginal posteriors. Each marginal posterior $\{q_{d}(z_{d}|\mathbf{m}, \mathbf{w})\}_{d = 1}^{D}$ is described as:
\begin{equation}
\begin{gathered}
 q_{d}(z_{d}|\mathbf{m}, \mathbf{w}) \sim \mathcal{G}^{\tau}(z_{d}| \mathbf{m}, \mathbf{w}, \mu_{d}, \sigma_{d}^{2}, \mathcal{A});\\
\mathcal{G}^{\tau}(z_{d}|\mathbf{m}, \mathbf{w}, \mu_{d}, \sigma_{d}^{2}) =  \frac{1}{c_{\text{tr}}}\mathcal{G}(z_{d}|\mathbf{m}, \mathbf{w}, \mu_{d}, \sigma_{d}^{2}, \mathcal{A}) \mathbb{I}(z_{d} \in \mathcal{A}),
\end{gathered}
\label{eq:Gaussian_marg_expression}
\end{equation}
where $\mathcal{G}^{\tau}$ denotes the truncated Gaussian distribution with mean $\mu_{d}$, variance $\sigma_{d}^{2}$, support region $\mathcal{A} = [b_{\text{low}}, b_{\text{up}}]^{D}$, and normalizing constant $c_{\text{tr}}$. 
We can thus express the final  joint PDF as: 
\begin{equation}
    q^{\text{Cop}}_{\bm{\theta}}(\mathbf{z}|\mathbf{m},\mathbf{w}) := \frac{1}{\sqrt{2\pi |\hat{\bm{\Sigma}}|}}\exp \left( - \frac{1}{2}\left( \mathbf{F}_{\text{TG}}^{-1}(\mathbf{u})\right)^{\top} (\hat{\bm{\Sigma}} ^{-1} - \mathbf{I})\mathbf{F}_{\text{TG}}^{-1}(\mathbf{u}) \right)\prod_{d=1}^{D} q_{d}(z_{d}|\mathbf{m}, \mathbf{w}),
    \label{eq:joint_gaussian_copula_density}
\end{equation}
where $F_{\text{TG}}(\cdot)$ and  $F_{\text{TG}}^{-1}(\cdot)$ denote the truncated Gaussian CDF and inverse CDF, respectively. 
We recall that the expression of the truncated Gaussian CDF is:
$F_{\text{TG}}(x_{i}) = \frac{\Phi\left(\frac{x_{i} - \mu_{i}}{\sigma_{i}}\right) - \Phi\left(\frac{a_{i} - \mu_{i}}{\sigma_{i}}\right)}{\Phi\left(\frac{b_{i} - \mu_{i}}{\sigma_{i}}\right) - \Phi\left(\frac{a_{i} - \mu_{i}}{\sigma_{i}}\right)}$, where $\Phi(x_{i}) = \int_{-\infty}^{x_{i}} \frac{1}{\sqrt{2\pi}}e^{-t^{2}/2} dt$ is the standard Gaussian CDF.

We parameterize the covariance matrix $\hat{\bm{\Sigma}}$ as $\hat{\bm{\Sigma}}=\hat{\mathbf{L}}\hat{\mathbf{L}}^{\top}$, where $\hat{\mathbf{L}}$ is a lower-triangular matrix. We include the entries of $\hat{\mathbf{L}}$ as part of the output of the DNN so that $\hat{\mathbf{L}}$ is a function of $\bm{\theta}$.
Thus, the parameters to be estimated by the encoder $\mathcal{E}_{\bm{\theta}}(\cdot)$ in this approach include the means and variances of the marginals, and the non-zero entries of the lower-triangular factorization matrix (denoted as $\hat{\mathbf{l}}$), i.e.: $\bm{\zeta}(\bm{\theta}) = \{ \bm{\mu}(\bm{\theta}), \bm{\sigma}(\bm{\theta}),\hat{\mathbf{l}}(\bm{\theta})\}$, where $\bm{\mu}(\bm{\theta})\in{\mathbb{R}^{D}}$, $\bm{\sigma}(\bm{\theta})\in{\mathbb{R}^{D}}$, and $\hat{\mathbf{l}}(\bm{\theta}) \in \mathbb{R}^{n_{L}}$ $\left(n_{L} = \frac{D(D-1)}{2}\right)$.


\textbf{Algorithm to sample from the copula PDF:}
\label{sec:sampling_copula}
for any input measurement $\{\mathbf{m}, \mathbf{w}\}$, the encoder $\mathcal{E}_{\bm{\theta}}([\mathbf{m}, \mathbf{w}])$ estimates the parameters $\bm{\zeta}(\bm{\theta})$ describing the posterior PDF.
To produce $H$ samples from this distribution, we first sample a joint zero-mean Gaussian vector with covariance matrix $\hat{\Sigma}(\bm{\theta}) = \hat{\mathbf{L}}(\bm{\theta})\hat{\mathbf{L}}(\bm{\theta})^{\top}$ by setting $\mathbf{w} = \hat{\mathbf{L}}(\bm{\theta})\mathbf{a}$, where $\mathbf{a}$ is a standard Gaussian vector. Next, we apply the copula transformation setting $z_{d} = \mu_{d}(\bm{\theta}) + \sigma_{d}(\bm{\theta}) \odot F^{-1}(\Phi(w_{d}))$, where $\mu_{d}(\bm{\theta})$ and $\sigma_{d}(\bm{\theta})$ are the estimated mean and standard deviation of $z_{d}$, and the operator $\odot$ denotes the entrywise vector product.

\subsection{Parametric dimensionality of the models and expected scalability}
A critical aspect is the impact of the number of parameters on the parameters to be estimated, which affects the problem's complexity and the computational effort required during training. 
As we increase the dimensionality of the problem (e.g., to model more complex damage conditions, with more components and damage types), using Gaussian mixtures becomes prohibitive due to the increase in the number of parameters to be estimated (which prevents reaching enough accuracy). 
Conversely, copula models are expected to be more scalable and more efficient.

\begin{table}[htbp]
\centering
\label{tab:parameters}
\begin{tabular}{@{}cccccc@{}}
\toprule
\multicolumn{1}{p{2.5cm}}{\centering \textbf{Model} } & \multicolumn{1}{p{5.5cm}}{\centering \textbf{Number of Parameters} } & \multicolumn{4}{c}{\textbf{Examples for $\mathbf{K=5}$}} \\
\cmidrule(lr){3-6}
 & & $D=2$ & $D=3$ & $D=5$ & $D=10$ \\
\midrule
$q_{\bm{\theta}}^{\text{GM}_{\text{diag}}}$ & $K(2D+1) - 1 $ & $24$ & $34$ & $54$ & $104$ \\
\midrule
$q_{\bm{\theta}}^{\text{GM}_{\text{full}}}$  & $K\left(1 + D + \frac{D(D+1)}{2}\right) - 1$ & $29$ & $49$ & $104$ & $329$ \\
\midrule
$q_{\bm{\theta}}^{\text{Cop}}$  & $2D + \frac{D(D-1)}{2}$ & $5$ & $9$ & $20$ & $65$ \\
\bottomrule
\end{tabular}
\caption{\textit{Number of parameters to estimate for different models. Examples assume $K=5$ for GMs.}}
\end{table}

\section{Implementation and training}
\label{Sec:NN_specifications}
This section includes the specifications taken to implement and train the proposed VAE architecture (introduced in Section~\ref{Sec:VAE_arch}) to solve the inverse problem of damage identification in an FOWT. 
We employ the software: TensorFlow (version 2.15) and TensorFlow Probability (version 0.23) to load and process the datasets, build the VAE model, and train it~\cite{tensorflow2015-whitepaper}.

\subsection{Data pre-processing}
\label{sec:data_preprocessing}
The available data generated according to Section~\ref{Sec:Data_generation} is used to train the VAE architecture.
We first split our dataset $\mathcal{D}$ into $\mathcal{D}^{\text{train}}$ with $N^{\text{train}}$ training samples, $\mathcal{D}^{\text{val}}$ with $N^{\text{val}}$ validation samples, and $\mathcal{D}^{\text{test}}$ with $N^{\text{test}}$ testing samples, where the number of samples corresponds to 70, 20, and 10\% of the total, respectively. Table~\ref{tab:dataset_split} summarizes the specifications of each dataset. 
We rescale the data based on the training set to balance the contribution of the involved features regardless of their magnitude. 
We use the MinMax scaler \cite{minmax} to constrain the measurements into the interval $[0,1]$. 
According to Section \ref{Sec:Data_generation}, there are six rigid body DOFs describing the platform's motion. However, not all of them will be available in practice. 
It is a common situation to count only on clinometers measuring rotation DOFs, such as roll.  
In this work, we assume only one measurable DOF, roll, for simplicity: the main objective here is to investigate which method is the most suitable for tackling the FOWT damage identification problem, and we will design methods tackling all DOFs in future work.
Given that we extracted five representative features to describe any time-domain response signals, we end up with a motion measurement vector $\mathbf{m}\in \mathbb{R}^{5}$, corresponding to the roll DOF. 
The external environmental conditions --- wind velocity, wave height, and wave peak --- form the excitation vector $\mathbf{w}\in \mathbb{R}^{3}$. 

\begin{table}[htbp]
    \centering
    \vspace{0.2cm}
    \begin{tabular}{lcc} 
        \toprule
        & \multicolumn{2}{c}{\textbf{Volume}} \\
        \cmidrule(lr){2-3}
        \textbf{Subset} & \textbf{Samples} & \textbf{Share} (\%) \\ 
        \midrule
        \textbf{Train} & 82,788 & 70.0  \\
        \textbf{Validation} & 23,653  & 20.0 \\
        \textbf{Test}    & 11,827  & 10.0 \\ 
        \midrule
        \textit{Total}      & \textit{118,268} & \textit{100.0} \\
        \bottomrule
    \end{tabular}
    \caption{\textit{Detailed distribution of the dataset across train, validation, and test subsets. The table includes sample counts and percentages relative to the total dataset.}}
\label{tab:dataset_split}
\end{table}

\subsection{VAE architecture layers}
According to Section~\ref{Sec:inverse_problem_formulation}, the VAE architecture contains three main components: (i) the encoder $\mathcal{E}_{\bm{\theta}}(\cdot)$, which maps the input measurements $\{\mathbf{m}, \mathbf{w}\}$ to the parameters $\bm{\zeta}(\bm{\theta})$ of the approximate posterior PDF $q_{\bm{\theta}}(\mathbf{z}|\mathbf{m},\mathbf{w})$; (ii) the sampling layer, which builds the approximate posterior PDF and generates samples $\mathbf{z}(\bm{\theta})$ from it; and (iii) the decoder $\mathcal{F}_{\bm{\varphi}^{*}}\left(\mathbf{z}(\bm{\theta}),\mathbf{w}\right)$, which approximates the exact forward operator $\mathcal{F}(\cdot)$. 
In Table~\ref{tab:arch_specs}, we summarize the architecture specifications, including layer dimension, activation function, and weight initialization method. 
\begin{table}[h!]
    \centering
    \begin{tabular}{c l c}
        \toprule
        \multicolumn{2}{l}{Encoder $\mathcal{E}_{\bm{\theta}}$} & \\
        \bottomrule
         & Layer units & 100, 250, 300, 300, 200, 150, 100\\
         & Activations & ReLU, ReLU, Tanh, ReLU, Tanh, ReLU, Tanh \\
         & Weight initialization & GU, GU, HU, GU, HU, GU, HU \\
        \midrule
        \multicolumn{2}{l}{Activations for encoder output layer } & \\
        \midrule
        & Means $\bm{\mu}$ & Sigmoid \\
        & Variances $\bm{\sigma}$ & Softplus \\
        & Lower triangular entries $\mathbf{l}$ and $\hat{\mathbf{l}}$& Linear \\
        
        & Weighting factors $\bm{\alpha}$ & Softmax \\
        \midrule
        \multicolumn{2}{l}{Decoder (trained in \cite{Navamuel_wes2025})} & \\
        \midrule
         & Layer units & 10, 30, 50, 70, 80\\
         & Activations & Tanh, ReLU, ReLU, ReLU, ReLU \\
         & Weight initialization & GU, HU, HU, HU, HU \\
         \bottomrule
    \end{tabular}
    \caption{\textit{Architecture specifications for the proposed VAE. ReLU: rectified linear unit; GU: Glorot uniform, HU: He uniform.}}
    \label{tab:arch_specs}
\end{table}

$\mathbf{Encoder}$: The encoder ($\mathcal{E}_{\bm{\theta}}$) is a fully connected NN with hidden layers, following the same architecture as work~\cite{Navamuel_wes2025}. 
This architecture combines two activation functions for the hidden layers: hyperbolic tangent (Tanh) \cite{ashkan2009efficient} and Rectified Linear Unit (ReLU) \cite{agarap2018deep}. We use Glorot and He uniform initializations for the hyperbolic tangent and ReLU layers, respectively~\cite{aldirany2024accurate}. 
For the output layer, we must consider different activation functions depending on the features to be estimated. 
We use the \textit{sigmoid} function \cite{han1995influence} to estimate means as it yields a smooth output in the interval $[0,1]$. 
We use the \textit{softmax} function \cite{goodfellow2016deep} for the weighting factors $\bm{\alpha}$ that build the Gaussian mixture, ensuring that their sum is equal to one and each value ranges into $[0,1]$.
For the non-zero entries of the lower-triangular factorization matrices $\mathbf{L}$ and $\hat{\mathbf{L}}$, we apply \textit{linear} activation. 

\textbf{Sampling layer}: The sampling layer builds the posterior density function $q_{\bm{\theta}} (\mathbf{z}|\mathbf{m}, \mathbf{w})$ from the estimated properties $\bm{\zeta}({\bm{\theta}})$ and draws $H$ samples $\mathbf{z}(\bm{\theta})$ that represent different damage conditions of the system. 
The specifications of the sampling layer depend on the selected posterior distribution model (see the details in sections \ref{sec:Gaussian_Mixture_posterior} and \ref{sec:Gaussian_Copula_posterior}).
It embeds the sampling algorithm within the VAE architecture, allowing it to be trained via automatic differentiation.
The drawn samples $\mathbf{z}(\bm{\theta})$ enter the decoder $\mathcal{F}_{\bm{\varphi}^{*}}$ together with the observed operating conditions in $\mathbf{w}$.  

\textbf{Decoder}: The decoder $\mathcal{F}_{\bm{\varphi}^{*}}$ is a fully-connected NN that maps the estimated system condition $\mathbf{z}(\bm{\theta})$ and the external excitation $\mathbf{w}$ to its motion response $\hat{\mathbf{m}}$. 
It also contains a set of hidden layers with ReLU and Tanh activation functions. The output layer applies a sigmoid activation function given the rescaling of the input measurements $\mathbf{m}$ to the interval $[0,1]$.


\subsection{Training}
Here we follow a similar two-step training strategy as in~\cite{Pardo_loss_inverse, GOROSTIDI2023115862, Navamuel_wes2025} (cf. Section~\ref{Sec:VAE_arch}), according to which the decoder is pretrained so that it approximates the forward operator $\mathcal{F}$ and then kept fixed in the second step~\cite{GOROSTIDI2023115862}.

In Section~\ref{Sec:ELBO_loss}, we introduced the ELBO function (see Eq.~\eqref{eq:ELBO_loss}) that accounts for the measurement misfit (likelihood term) and the uncertainty (posterior entropy term). 
Finding an optimal (or at least a sufficiently good) encoder $\mathcal{E}_{\bm{\theta}^{*}}(\cdot)$ demands optimizing a loss function $\mathcal{L}_{\text{ELBO}}$ based on the ELBO value.
Here, we approximate $\mathcal{L}_{\text{ELBO}}$ as an average over the training dataset ($\mathcal{D}^{\text{train}} = \{\mathbf{m}_{i}, \mathbf{w}_{i}\}_{i=1}^{N^{\text{train}}}$) with $H$ samples drawn from the approximate posterior PDF at each iteration, yielding the empirical loss function:
\begin{equation}
\mathcal{L}_{\text{ELBO}}(\bm{\theta})\approx \frac{1}{N^{\text{train}} \cdot H} \sum_{i = 1}^{N^{\text{train}}}\sum_{h=1}^{H}   \left[\frac{1}{2} (\mathbf{m}_{i}  -\mathcal{F}_{\bm{\varphi}^{*}}([ \mathbf{z}_{i}^{h},\mathbf{w}_{i}]))^{\top} \Gamma^{-1} ( \mathbf{m}_{i} - \mathcal{F}_{\bm{\varphi}^{*}}([\mathbf{z}_{i}^{h},\mathbf{w}_{i}]))  + \text{log} \ q_{\bm{\theta}}( \mathbf{z}_{i}^{h}|\mathbf{m}_{i},\mathbf{w}_{i}) \right],
\label{eq:ELBO_minimization}
\end{equation}
where $\Gamma = \text{diag}(\beta^{2})$ with $\beta = 0.075$, corresponding to noise levels of up to $\approx 8.5\%$~\cite{bishop2006pattern} (values determined empirically in prior work~\cite{Navamuel_wes2025}). 
The number of samples to draw is set to $H = 1$, as it is sufficient to ensure convergence, and a larger number of samples would imply more time-consuming iterations. 
Note the sign change with respect to the expression for the ELBO in Eq.\eqref{eq:ELBO_loss} as we want to maximize the ELBO during training. 


Obtaining the optimal parameter set $\bm{\theta}^{*}$ can thus be expressed as a minimization problem: 
\begin{equation}
\bm{\theta}^{*}:= \argmin_{\bm{\theta}}\mathcal{L}_{\text{ELBO}}(\bm{\theta};\mathcal{D}^{\text{train}}).
\label{eq:ELBO_minimization}
\end{equation}

We implement a gradient-based optimization solver in Tensorflow 2.15 using the Adam optimizer~\cite{Kingma2015_Adam}. 
We set the initial learning rate (LR) to $10^{-4}$. This value was obtained via a trial-and-error process to balance convergence speed and loss value instability. 
The batch size (number of samples from the training set $\mathcal{D}^{\text{train}}$ tackled at each iteration before the DNN parameters ($\bm{\theta}$) are updated) is set to 1,024. As a stopping criterion, we terminate the training process when the loss value does not change by more than $ 10^{-3}$ across 1,000 epochs in absolute terms.

\section{Numerical results}
This section describes the numerical results obtained for the three approaches considered to approximate the true posterior PDF $p(\mathbf{z}|\mathbf{m}, \mathbf{w})$, namely a Gaussian mixture with diagonal covariance ($q_{\bm{\theta}}^{\text{GM}_{\text{diag}}}(\mathbf{m},\mathbf{w})$), a Gaussian mixture with full covariance ($q_{\bm{\theta}}^{\text{GM}_{\text{full}}}(\mathbf{m}, \mathbf{w})$), and a Gaussian copula with Gaussian marginals ($q_{\bm{\theta}}^{\text{Cop}}(\mathbf{m},\mathbf{w})$).
We first compare the training cost and performance over the test dataset $\mathcal{D}^{\text{test}}$, which contains $N^{\text{test}}$ samples --- unseen during training --- corresponding to $10\%$ of the available data (see Section~\ref{sec:data_preprocessing}). 
Next, we study the variability in the ELBO value for the unseen test samples. 
Finally, we represent the approximate posterior PDF and compare it with the true posterior (ground truth) obtained via randomized quasi-Monte Carlo (QMC) sampling.

\subsection{Comparison of computational cost and test performance}
In this section, we compare the computational effort required by each posterior approximation method, as well as the achieved testing performance. 
For the Gaussian mixture models ($q_{\bm{\theta}}^{\text{GM}_{\text{diag}}}$ and $q_{\bm{\theta}}^{\text{GM}_{\text{full}}}$), we consider an increasing number of components, i.e., $\text{K} = \{1, 2, 5, 10\}$.
In Table~\ref{tab:metrics_comparison}, we compare the nine approaches according to four relevant metrics: the number of distributional parameters to be estimated, the training time, the average ELBO value over the test dataset ($\text{Avg}. \ \text{ELBO}^\text{test}$, and the Bayesian Information Criterion (BIC) metric~\cite{joe1996estimation}. 
The number of parameters and training time are used as metrics for model complexity and computational cost, while the ELBO values measure the generalization capability of the method after training.
The BIC provides a hybrid metric that penalizes the model complexity against the attained accuracy. 
For any probabilistic model parameterized by $p$ parameters ($\bm{\zeta}(\bm{\theta}) \in \mathbb{R}^{p}$) we calculate the BIC as:
\begin{equation}
    \text{BIC} = -2 \frac{\hat{L}}{N} + \frac{p \log(N)}{N},
\label{eq:BIC_metric}
\end{equation}
where $\hat{L}_{i}$ denotes the model log likelihood (one of the terms forming the ELBO, see Eq.\eqref{eq:ELBO_derivation}), $N$ indicates the number of points in the dataset, and $p$ is the number of parameters of the chosen PDF model. 
Note that the smaller the BIC, the better.
The first term in Eq.~\eqref{eq:BIC_metric} reflects how well the model fits the data, while the second term penalizes the model's complexity, measured by the number of parameters (the more parameters, the higher the cost and the more difficult the optimization).

\begin{table}[h!]
\centering 
\begin{tabular}{l cccc cccc c} 
\toprule
& \multicolumn{4}{c}{${\text{GM}_{\text{diag}}}$}
& \multicolumn{4}{c}{$\text{GM}_{\text{full}}$}
& \multicolumn{1}{c}{Copula} \\ 
\cmidrule(lr){2-5} \cmidrule(lr){6-9} \cmidrule(lr){10-10} 
\textbf{Metric} & K = 1 & K = 2 & K = 5&  K = 10 & K = 1 & K = 2 & K = 5 & K = 10 &  \\ 
\midrule %
Parameters & 4 & 9 & 24 & 29 & 5 & 10 & 29 &  50 & 5 \\
Train time (h) & 3.36 & 4.95 & 3.17 & 3.70 & 6.56 & 7.95 &  7.15 & 7.18 &  3.32 \\
Avg. $\text{ELBO}^{\text{test}}$ & 6.07 & 6.22 &  6.24 & 6.27 & 6.98 & 6.99 & 6.95 &  6.87 & 7.03 \\
BIC & -15,887 &-16,024 & -15,958 & -15,810 & -18,145 & -18,115 & -17,853&   -17,499 & -18,213 \\
\bottomrule 
\end{tabular}
\caption{\textit{Performance comparison for different posterior models}} 
\label{tab:metrics_comparison}
\end{table}
Results are shown in Table 4. As expected, the computational cost increases with the number of model parameters.
We note that the Gaussian mixture approaches are more expensive. We believe this is due to both the higher number of parameters and the increased computational cost arising from truncating a D-dimensional distribution and employing rejection sampling. The copula approach only truncates one-dimensional marginals and employs fewer parameters, resulting in higher computational efficiency.
Not only is the copula strategy fast, but it is also accurate. 
Indeed, it achieves comparable accuracy metrics to the Gaussian mixture models with full covariances. 
In contrast, the Gaussian mixture models with diagonal covariances are consistently less accurate.

These results reveal the need for a posterior distribution that captures correlations between the damage condition features in $\mathbf{z}$, which is hardly attainable in the diagonal covariance case. We observe that as the number of components increases, the average ELBO value for the test dataset (Avg. $\text{ELBO}^{\text{test}}$) increases slightly. 
However, this increase is counterbalanced by an increase in model complexity, which may hinder training. 
Considering a full covariance enhances the results even with one single component, as it can better represent correlations. 
We note that negligible improvements are achieved as $K$ increases. We suspect that for large $K$, the parameters are too many, making the optimizer struggle and converge to a local minimum. We leave the investigation of techniques for ameliorating this behavior to future work.

The BIC values in Table~\ref{tab:metrics_comparison} reflect a similar behavior: for the mixture models, the BIC actually worsens as $K$ is increased, suggesting that the additional parameters introduce unnecessary complexity without yielding meaningful gains. On the other hand, the copula model achieves the best BIC value, representing a good balance between model complexity and accuracy.
While the single-component full covariance model ($\text{GM}^{\text{full}},\ K=1$) provides a competitive result, the copula achieves the optimal balance between model complexity and accuracy. 
The BIC values for the full covariance approaches worsen (become less negative) as $K$ increases from 1 to 10 (from $-18,145$ to $-17,499$), indicating that the additional parameters introduce unnecessary complexity without meaningful gain.

\subsection{ELBO variability across the test dataset}
\label{sec:stochasticity_effect}
We now investigate how the ELBO varies across the different scenarios described by our test set measurements.
Informed by the results from Table~\ref{tab:metrics_comparison}, we select the best-performing model for each type (diagonal covariance, full covariance, copula): the Gaussian mixture with diagonal covariance and $K = 10$ components, the Gaussian mixture with full covariance and $K = 2$ components, and the Gaussian copula model.

\begin{figure}[h!]
    \centering
    \includegraphics[width=0.72\linewidth]{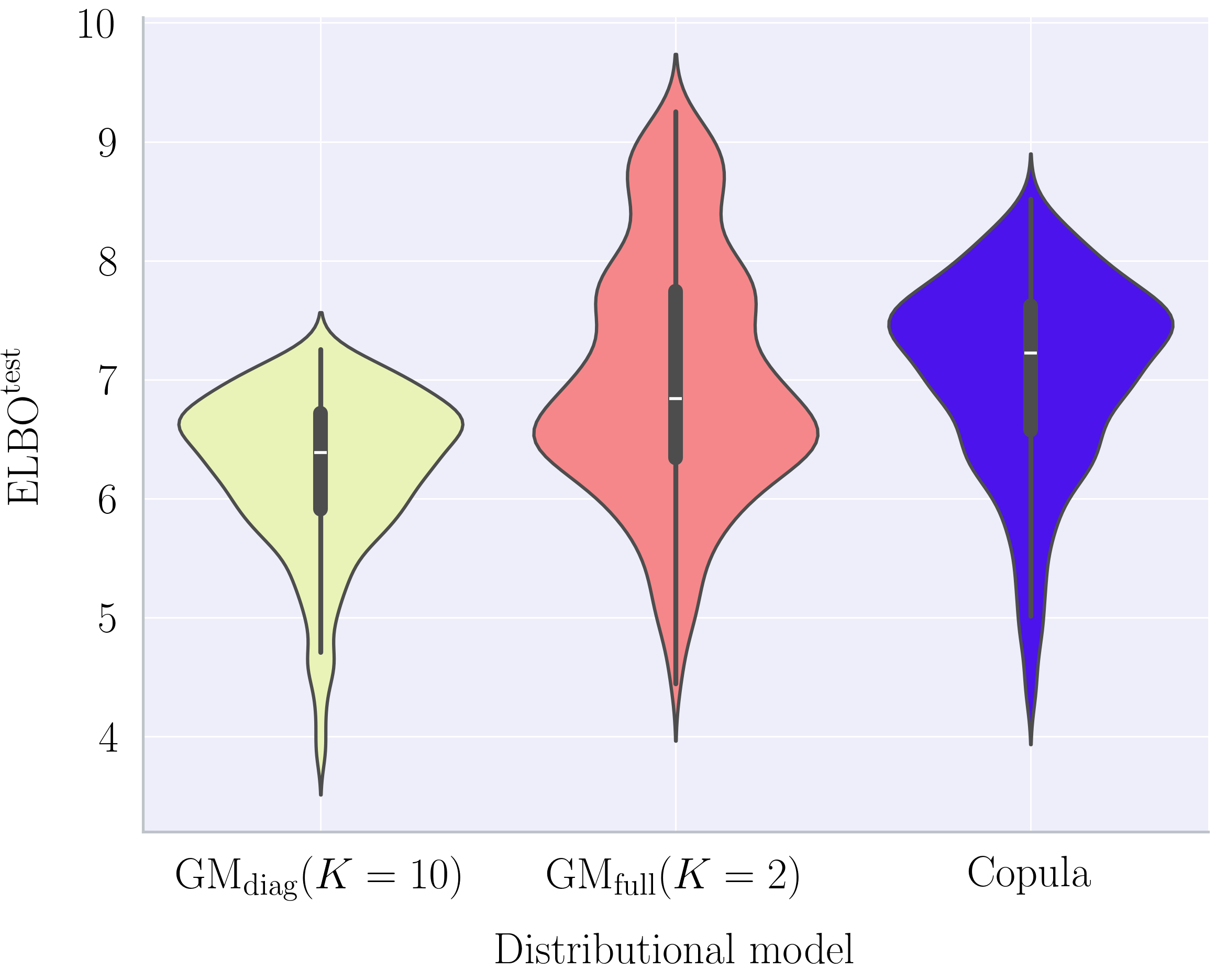}
    \caption{\textit{Violin diagrams of the $\text{ELBO}$ value for different posterior models, calculated across $\mathcal{D}^{\text{test}}$}}
    \label{fig:violin_test_ELBO_plot}
\end{figure}

In Figure~\ref{fig:violin_test_ELBO_plot}, we represent the violin diagrams of the ELBO for the test dataset ($\text{ELBO}^{\text{test}}$) for the three selected approaches. 
The results highlight distinct performance characteristics depending on the expressiveness of each posterior parameterization. 
The diagonal covariance Gaussian mixture model (left, light green shade) exhibits a pronounced lower tail extending to negative ELBO values, indicating poor performance in a subset of test scenarios. 
This behavior is expected since the restriction to a diagonal covariance matrix prevents the model from capturing the existing correlations between the features in $\mathbf{z}$. 
Consequently, even when increasing the number of components ($K$), the model struggles to accurately approximate the posterior when strong dependencies exist, resulting in the observed heavy lower tail.

In contrast, the full covariance Gaussian mixture model (center, pink shade) improves robustness. 
The lower tail is shorter compared to the diagonal case, and the distribution extends further upwards, reflecting higher ELBO values for some test scenarios. 
This behavior aligns with our expectations: employing a full covariance improves the expressiveness of the posterior, as it enables capturing the existing correlations.

Finally, the copula approximation (right, blue shade) achieves the highest median (indicated by a horizontal white mark), demonstrating consistently accurate approximations. 
However, the copula violin plot also exhibits a long lower tail corresponding to test data with multimodal true posteriors. 
Indeed, while copulas efficiently capture the correlations between damage conditions, they are also inherently unimodal and --- unlike mixtures --- cannot capture multimodality.

\subsection{Posterior PDF comparison}
This section compares the approximate posterior PDF plots for specific test scenarios. 
We approximate the true posterior (ground truth) using quasi-Monte Carlo (QMC) sampling to establish a baseline for comparison. 

\subsubsection{Ground truth definition}
 \label{sec:ground_truth_def}
To provide a visual comparison between the true and the estimated posterior distributions for any measured input data $\{\mathbf{m}, \mathbf{w}\}$, we need to have access to the true posterior. 
According to the proportionality in Eq.~\eqref{eq:post_bayes}, we can express the posterior as: 
\begin{equation}
    p(\mathbf{z}|\mathbf{m}, \mathbf{w}) = \frac{1}{\text{B}}p(\mathbf{m}|\mathbf{z},\mathbf{w})\cdot p(\mathbf{z}),
\label{eq:proportionality_posterior}
\end{equation}
where $p(\mathbf{z}) = \frac{\mathds{1}_{\mathcal{A}}(\mathbf{z})}{\text{vol}(\mathcal{A})}$ corresponds to the uniform PDF in the finite domain $\mathcal{A}$, and B acts as the (inverse) proportionality constant.

By embedding the likelihood expression of Eq.~\eqref{eq:conditional_gaussianoise} into  Eq.~\eqref{eq:proportionality_posterior}, we can estimate $\text{B}$ using the PDF properties as: 
\begin{equation}
\text{B} = \text{B} \int_{\mathcal{A}}p(\mathbf{z}|\mathbf{m},\mathbf{w})dz = \int_{\mathcal{A}}\underbrace{\frac{\mathds{1}_{\mathcal{A}}(\mathbf{z})}{\text{vol}(\mathcal{A})}}_{\textcolor{blue}{p(\mathbf{z}})} \underbrace{\frac{1}{(2\pi)^{M/2}|\Gamma|^{1/2}}\cdot \exp\left(- \frac{1}{2} (\mathbf{m}-\hat{\mathbf{m}}(\mathbf{z})^{\top}\Gamma^{-1}(\mathbf{m} - \hat{\mathbf{m}}(\mathbf{z})) \right)}_{\textcolor{blue}{p(\mathbf{m}|\mathbf{z},\mathbf{w})}}dz,
\end{equation}
where, for simplicity in notation, we denote $\hat{\mathbf{m}}(\mathbf{z})$ to the decoder output: $\hat{\mathbf{m}}(\mathbf{z}) = \mathcal{F}_{\bm{\varphi}^{*}}([\mathbf{z},\mathbf{w}])$. 
We employ the randomized quasi Monte Carlo (QMC)~\cite{owen2000monte, l2002recent} to approximate this integral: 
\begin{equation}
    \text{B} = \int_{\mathcal{A}} p(\mathbf{z}|\mathbf{m},\mathbf{w})dz \approx \frac{1}{\text{N}_{\text{QMC}}}\sum_{j =1}^{\text{N}_{\text{QMC}}}p(\mathbf{z}_{j}|\mathbf{m},\mathbf{w}),
\end{equation}
where $\text{N}_{\text{QMC}}$ represents the number of points generated via randomized QMC sampling.
In this work, we consider $\text{N}^{\text{QMC}} =2^{14} = 16,384$ QMC points.

Once the constant $\text{B}$ is obtained, we can evaluate the true posterior for a set of $\text{N}_{\text{plot}}$ points and plot the PDF.
This plot describes the damage conditions $\mathbf{z} = \{z_{1}, z_{2}\}$ that are most likely to have produced the motion response $\mathbf{m}$ given the external conditions $\mathbf{w}$.
For illustrative purposes, we employ this strategy to approximate the true posterior in four test scenarios randomly selected from $\mathcal{D}^{\text{test}}$. 
We remark that, generally speaking, $B$ is computationally intractable, and it is only possible to approximate it here since we work in relatively low dimensions. 
However, its computation enables us to construct the ground truth, which we can then use as a comparison to validate our methods.

\subsubsection{Comparison for four test scenarios}
For the three selected models (see Figure~\ref{fig:violin_test_ELBO_plot}), we compare the resulting approximated posterior PDFs with the ground truth (computed via the techniques just described in  Section~\ref{sec:ground_truth_def}) for four test scenarios (randomly drawn from $\mathcal{D}^{\text{test}}$).
\begin{figure}[h!]
    \centering
    \includegraphics[width=0.99\linewidth]{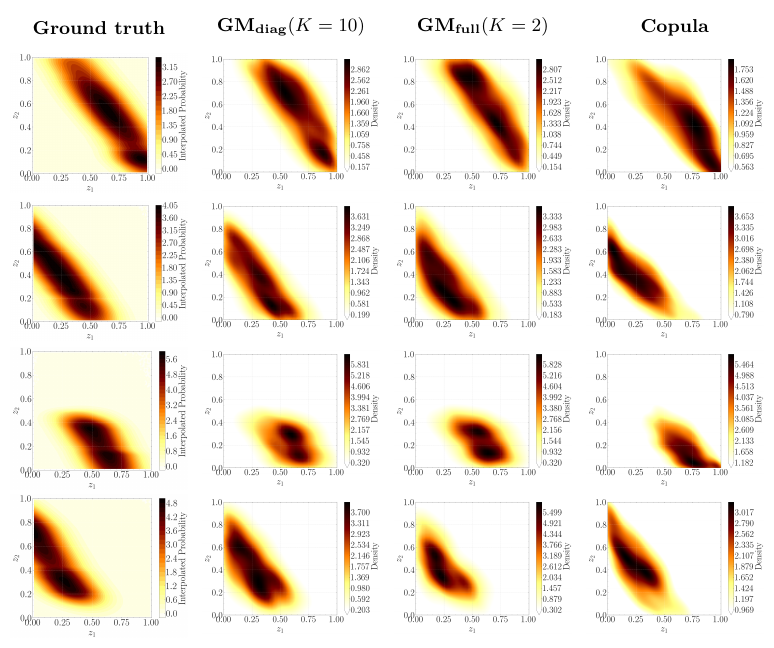}
    \caption{\textit{Comparison of three approaches against ground truth for four illustrative test scenarios. Each row corresponds to one specific scenario and contains, from left to right: ground truth, Gaussian mixture with diagonal covariance and $K = 10$ components, Gaussian mixture with full covariance and $K = 2$ components, and Gaussian copula.}}
    \label{fig:KDE_comparison}
\end{figure}
The ground truth figures reveal the correlation structure between the two damage conditions. This solution is challenging to achieve with a diagonal covariance Gaussian mixture PDF ($q_{\bm{\theta}}^{\text{GM}_{\text{diag}}}$), where the variables are assumed to be uncorrelated. 
As we stack many components ($K = 10$), we are able to capture this correlation at the cost of increasing the model complexity. 
The figure shows that the full covariance Gaussian mixture ($q_{\bm{\theta}}^{\text{GM}_{\text{full}}}$ successfully captures the target solutions even with a reduced number of components ($K = 2$). 
The copula approach also matches the ground truth representation, adopting more irregular shapes due to the superimposing effect of the multivariate standard Gaussian over the truncated Gaussian marginals. However, this approach remains restrictive, as it prevents the representation of multimodal solutions.

\section{Conclusions and future work}
\textbf{Paper overview:} This paper constructed and compared different probabilistic models for approximating the posterior distribution of a VAE architecture designed to identify damage in FOWT mooring systems. 
We assessed three distinct approaches: a truncated GM model with a diagonal covariance matrix, a truncated GM model with a full covariance matrix, and a Gaussian copula model with truncated Gaussian marginals. 
We compared the performance of the three approaches by assessing their testing performance and the computational cost of training. 

Numerical experimentation demonstrates that the copula approach provides an efficient methodology for solving the two-dimensional problem of identifying biofouling and anchoring damage.
Indeed, it achieved superior performance on the test set, using fewer parameters, compared to the GM approaches. 
These results suggest that copula-based VAEs offer a scalable methodology for tackling higher-dimensional problems
The analysis conducted in Section~\ref{sec:stochasticity_effect} confirms the limitations of diagonal covariance approaches, which are unable to capture correlations.
In contrast, the full covariance and copula models enhance prediction accuracy, suggesting that accounting for correlations is crucial for identifying FOWT damage.
Additionally, the copula approach appears to be more robust, as its distribution is tightly concentrated around the optimal performance region, with shorter lower tails and closer outliers. These observations suggest that copula-based VAEs offer a robust and efficient technology for FOWT damage assessment.



Nevertheless, we acknowledge that this study is subject to a few limitations that call for future research. 
Currently, we identify four of them: 1) We use synthetic data generated via OpenFAST. 
While OpenFAST has been validated with extensive experimental data and is a well-established and reliable high-fidelity realistic data simulator~\cite{coulling2013validation, reig2024efficient}, the lack of experimental data from an operating FOWT implies that the proposed VAEs have
not been tested against real-world complexities (e.g., non-Gaussian noise, unmodeled dynamics, sensor drift, or unexpected environmental effects).
2) We have only taken into account static damage scenarios, which neglect the time-domain evolution of the system's condition, potentially hindering a long-term degradation analysis, including fatigue analysis and remaining useful-life estimates.
3) While our method effectively represents the multi-source uncertainty in the predictions, it is currently unable to disentangle the distinct contributions from sources such as measurement error, modeling error, and sparse sensing.
4) Despite its computational efficiency, the expressivity of the copula posterior PDF struggles at capturing complex, multimodal solutions.

\textbf{Future work:} A natural first step is to address the above-mentioned limitations. Firstly, we aim to incorporate real data from an operating FOWT to complement and enhance the existing synthetic training dataset. Secondly, we aim to develop models that can estimate the evolution of damage over time in conjunction with external loading. This estimation is crucial for fatigue analysis and estimating the Remaining Useful Life (RUL) of mooring systems. Subsequently, we aim to investigate techniques for disentangling the various sources of uncertainty based on their distinct behaviors~\cite{kamariotis2025Uncertainties}.
Lastly, we aim to enhance the expressiveness of the approximate posterior by equipping the copula with multimodal marginals, thereby providing a flexible, yet efficient, solution. 

Additionally, we plan to apply the proposed methodology to a dataset with a higher-dimensional latent space, thereby fully demonstrating its scalability. 
Moreover, we aim to further validate the proposed method by accounting for more realistic noise patterns, such as colored noise or impulse noise (to simulate sensor faults).
Future research will focus on optimizing instrumentation systems and sensor placement.

\section*{Acknowledgements}
Ana Fernandez-Navamuel has received support from the Juan de la Cierva Postdoctoral Fellowship under the Grant JDC2023-051132-I funded by MICIU/AEI/10.13039/501100011033 and by the FSE+;  by the European Union’s Horizon Europe research and innovation programme under Grant Agreement 101162248 -ORE4Citizens; by the Elkartek program under Grants KK-2024/00068 (SEGURH2) and KK-2024/00086 (RUL-ET); by the IKUR-HPC$\&$AI program (HPCAI7.OceaNNic); and by the Basque Government through the BERC 2022-2025 program and by the Ministry of Science and Innovation: BCAM Severo Ochoa accreditation CEX2021-001142-S / MICIN / AEI / 10.13039/501100011033.

\noindent Martín A. Díaz-Viera acknowledges the support of the Mexican Petroleum Institute.

\noindent Matteo Croci has received support from the grant PID2023-146668OA-I00 funded by MICIU / AEI / 10.13039 / 501100011033 and cofunded by the European Union and by grant RYC2022-036312-I funded by MICIU / AEI / 10.13039 / 501100011033 and by ESF+. M. Croci is also supported by the Basque Government through the BERC 2022-2025 program, and by the Ministry of Science and Innovation: BCAM Severo Ochoa accreditation  CEX2021-001142-S / MICIN / AEI / 10.13039 / 501100011033

\pagebreak

\bibliographystyle{elsarticle-num}
\bibliography{references.bib}

@article{martinez2016structural,
  title={Structural health monitoring of offshore wind turbines: A review through the Statistical Pattern Recognition Paradigm},
  author={Martinez-Luengo, Maria and Kolios, Athanasios and Wang, Lin},
  journal={Renewable and Sustainable Energy Reviews},
  volume={64},
  pages={91--105},
  year={2016},
  publisher={Elsevier},
  doi = {10.1016/j.rser.2016.05.085},
}

@book{bishop2006pattern,
  title        = {Pattern Recognition and Machine Learning},
  author       = {Christopher M. Bishop},
  year         = {2006},
  publisher    = {Springer},
  address      = {New York},
  isbn         = {9780387310732},
  note         = {{I}SBN 9780387310732}
}

@misc{minmax,
author={\normalfont{Scikit Learn}},
title={MinMaxScaler Documentation},
year={2024},
url={https://scikit-learn.org/stable/modules/generated/ \ sklearn.preprocessing.MinMaxScaler.html},
note={Online; accessed 02-June-2024}
}

@misc{tensorflow2015-whitepaper,
title={ {TensorFlow}: Large-Scale Machine Learning on Heterogeneous Systems},
url={https://www.tensorflow.org/},
note={Software available from tensorflow.org},
author={
    Mart\'{i}n~Abadi and
    Ashish~Agarwal and
    Paul~Barham and
    Eugene~Brevdo and
    Zhifeng~Chen and
    Craig~Citro and
    Greg~S.~Corrado and
    Andy~Davis and
    Jeffrey~Dean and
    Matthieu~Devin and
    Sanjay~Ghemawat and
    Ian~Goodfellow and
    Andrew~Harp and
    Geoffrey~Irving and
    Michael~Isard and
    Yangqing Jia and
    Rafal~Jozefowicz and
    Lukasz~Kaiser and
    Manjunath~Kudlur and
    Josh~Levenberg and
    Dandelion~Man\'{e} and
    Rajat~Monga and
    Sherry~Moore and
    Derek~Murray and
    Chris~Olah and
    Mike~Schuster and
    Jonathon~Shlens and
    Benoit~Steiner and
    Ilya~Sutskever and
    Kunal~Talwar and
    Paul~Tucker and
    Vincent~Vanhoucke and
    Vijay~Vasudevan and
    Fernanda~Vi\'{e}gas and
    Oriol~Vinyals and
    Pete~Warden and
    Martin~Wattenberg and
    Martin~Wicke and
    Yuan~Yu and
    Xiaoqiang~Zheng},
  year={2015},
}

@phdthesis{aldirany2024accurate,
  title        = {Accurate Approximations of the Wave Equation:
From Spectral Element Methods to Deep Learning Approaches},
  author       = {Ziad Aldirany},
  year         = 2024,
  month        = {March},
  address      = {Montreal, Canada},
  school       = {University of Montreal},
  type         = {PhD thesis},
  URL          = {https://publications.polymtl.ca/58312/ }
}

@inproceedings{han1995influence,
  title={The influence of the sigmoid function parameters on the speed of backpropagation learning},
  author={Han, Jun and Moraga, Claudio},
  booktitle={International workshop on artificial neural networks},
  pages={195--201},
  year={1995},
  doi = {10.1007/3-540-59497-3_175},
  organization={Springer}
}

@misc{agarap2018deep,
      title={Deep {L}earning using Rectified Linear Units ({R}e{LU})}, 
      author={Abien Fred Agarap},
      year={2019},
      eprint={1803.08375},
      archivePrefix={arXiv},
      primaryClass={cs.NE},
      url={https://arxiv.org/abs/1803.08375}, 
}

@inproceedings{ashkan2009efficient,
author = {Namin, Ashkan and Leboeuf, Karl and Muscedere, Roberto and Wu, Huapeng and Ahmadi, Majid},
year = {2009},
month = {06},
pages = {2117 - 2120},
title = {Efficient hardware implementation of the hyperbolic tangent sigmoid function},
booktitle = {Proceedings - IEEE International Symposium on Circuits and Systems},
doi = {10.1109/ISCAS.2009.5118213}
}

@article{Mylonas2021,
author = {Mylonas, Charilaos and Abdallah, Imad and Chatzi, Eleni},
doi = {10.1002/WE.2621},
issn = {10991824},
journal = {Wind Energy},
month = {oct},
number = {10},
pages = {1122--1139},
publisher = {John Wiley and Sons Ltd},
title = {{Conditional variational autoencoders for probabilistic wind turbine blade fatigue estimation using Supervisory, Control, and Data Acquisition data}},
volume = {24},
year = {2021}
}

@article{Copulas_intro1,
author = {Schmidt, Thorsten},
year = {2007},
month = {01},
pages = {},
title = {Coping with copulas},
url={https://api.semanticscholar.org/CorpusID:15151026},
journal = {Copulas - From Theory to Application in Finance}
}

@article{GOROSTIDI2023115862,
title = {Diagnosis of the health status of mooring systems for floating offshore wind turbines using autoencoders},
journal = {Ocean Engineering},
volume = {287},
pages = {115862},
year = {2023},
issn = {0029-8018},
doi = {10.1016/j.oceaneng.2023.115862},
author = {N. Gorostidi and D. Pardo and V. Nava},
}

@article{Rodriguez2023,
author = {Rodriguez, Oscar and Taylor, Jamie M. and Pardo, David},
doi = {10.1093/gji/ggad362},
issn = {1365246X},
journal = {Geophysical Journal International},
number = {3},
pages = {2598--2613},
title = {{Multimodal variational autoencoder for inverse problems in geophysics: Application to a 1-D magnetotelluric problem}},
volume = {235},
year = {2023}
}

@InProceedings{BayesInverse_Goh2021,
  title = 	 {Solving Bayesian Inverse Problems via Variational Autoencoders},
  author =       {Goh, Hwan and Sheriffdeen, Sheroze and Wittmer, Jonathan and Bui-Thanh, Tan},
  booktitle = 	 {Proceedings of the 2nd Mathematical and Scientific Machine Learning Conference},
  pages = 	 {386--425},
  year = 	 {2022},
  editor = 	 {Bruna, Joan and Hesthaven, Jan and Zdeborova, Lenka},
  volume = 	 {145},
  series = 	 {Proceedings of Machine Learning Research},
  month = 	 {16--19 Aug},
  publisher =    {PMLR},
  url = 	 {https://proceedings.mlr.press/v145/goh22a.html},
}

@misc{openfast,
  author = "{OpenFAST Documentation}",
  title = "{OpenFAST Documentation}",
  url = "{https://openfast.readthedocs.io/en/main/}",
  note = "{Accessed: April 4, 2024}"
}

@article{VInference_ELBO2017,
archivePrefix = {arXiv},
arxivId = {1601.00670},
author = {Blei, David M. and Kucukelbir, Alp and McAuliffe, Jon D.},
doi = {10.1080/01621459.2017.1285773},
eprint = {1601.00670},
issn = {1537274X},
journal = {Journal of the American Statistical Association},
number = {518},
pages = {859--877},
title = {{Variational Inference: A Review for Statisticians}},
volume = {112},
year = {2017}
}

@article{Pardo_loss_inverse,
author = {Shahriari, Mostafa and Pardo, David and Rivera, Jon A. and Torres-Verdín, Carlos and Picon, Artzai and Del Ser, Javier and Ossandón, Sebastian and Calo, Victor M.},
title = {Error control and loss functions for the deep learning inversion of borehole resistivity measurements},
journal = {International Journal for Numerical Methods in Engineering},
volume = {122},
number = {6},
pages = {1629-1657},
doi = {10.1002/nme.6593},
year = {2021}
}

@book{faltinsen1993sea,
  title     = {Sea Loads on Ships and Offshore Structures},
  author    = {Faltinsen, Odd M.},
  series    = {Cambridge Ocean Technology Series},
  volume    = {1},
  year      = {1993},
  publisher = {Cambridge University Press},
  address   = {Cambridge},
  isbn      = {0-521-45870-6},
  note      = {{I}SBN 0-521-45870-6}
}

@article{jonkman2011dynamics,
author = {Jonkman, J. M. and Matha, D.},
title = {Dynamics of offshore floating wind turbines—analysis of three concepts},
journal = {Wind Energy},
volume = {14},
number = {4},
pages = {557-569},
doi = {10.1002/we.442},
year = {2011}
}

@article{NEWMAN1979221,
  title={The Theory of Ship Motions},
  author={John Nicholas Newman},
  journal={Advances in Applied Mechanics},
  year={1979},
  volume={18},
  pages={221-283},
  url={https://api.semanticscholar.org/CorpusID:117346186}
}

@article{Hatecke01062015,
author = {H. Hatecke},
title = {The impulse response fitting and ship motions},
journal = {Ship Technology Research},
volume = {62},
number = {2},
pages = {97--106},
year = {2015},
publisher = {Taylor \& Francis},
doi = {10.1179/2056711115Y.0000000001},
}

@article{jonkman2007dynamics,
author = {Jonkman, Jason M.},
title = {Dynamics of offshore floating wind turbines—model development and verification},
journal = {Wind Energy},
volume = {12},
number = {5},
pages = {459-492},
doi = {10.1002/we.347},
year = {2009}
}

@incollection{THEODORIDIS202019,
title = {Chapter 2 - Probability and Stochastic Processes},
editor = {Sergios Theodoridis},
booktitle = {Machine Learning (Second Edition)},
publisher = {Academic Press},
edition = {Second Edition},
pages = {19-65},
year = {2020},
isbn = {978-0-12-818803-3},
note = {{I}SBN 978-0-12-818803-3}, 
doi = {10.1016/B978-0-12-818803-3.00011-8},
author = {Sergios Theodoridis},
}

@article{Kingma2015_Adam,
archivePrefix = {arXiv},
arxivId = {1412.6980},
author = {Kingma, Diederik P. and Ba, Jimmy Lei},
journal = {3rd International Conference on Learning Representations, ICLR 2015 - Conference Track Proceedings},
pages = {1--15},
title = {{Adam: A method for stochastic optimization}},
year = {2015},
url={https://arxiv.org/abs/1412.6980}}

@misc{VAE_tutorial,
author = {Doersch, Carl},
year = {2016},
pages = {ArXiv Preprint},
title = {Tutorial on Variational Autoencoders},
doi = {https://doi.org/10.48550/arXiv.1606.05908},
}

@article{KL_Divergence,
 ISSN = {00034851},
 URL = {http://www.jstor.org/stable/2236703},
 author = {S. Kullback and R. A. Leibler},
 journal = {The Annals of Mathematical Statistics},
 number = {1},
 pages = {79--86},
 publisher = {Institute of Mathematical Statistics},
 title = {On Information and Sufficiency},
 urldate = {2025-06-25},
 volume = {22},
 year = {1951}
}

@article{Gibbs03102022,
author = {David Gibbs and Paul D. Jensen},
title = {Chasing after the wind? Green economy strategies, path creation and transitions in the offshore wind industry},
journal = {Regional Studies},
volume = {56},
number = {10},
pages = {1671--1682},
year = {2022},
publisher = {RSA Website},
doi = {10.1080/00343404.2021.2000958}}

@misc{gwec2025,
  author       = {GWEC}, 
  title        = {Global Wind Energy Report 2025},
  institution  = {Global Wind Energy Council},
  year         = 2025,
  address      = {Brussels, Belgium},
}

@article{JAHANI2022111136,
title = {Structural dynamics of offshore Wind Turbines: A review},
journal = {Ocean Engineering},
volume = {251},
pages = {111136},
year = {2022},
issn = {0029-8018},
doi = {10.1016/j.oceaneng.2022.111136},
author = {Kamal Jahani and Robert G. Langlois and Fred F. Afagh}
}

@article{Milligan_Workers1014,
author = {Gemma Milligan and Joseph O’Halloran and Mike Tipton},
title ={Quantifying the essential tasks of offshore wind technicians},
journal = {WORK},
volume = {77},
number = {4},
pages = {1245-1259},
year = {2024},
doi = {10.3233/WOR-230267}}

@article{KHAZAEE20221568,
title = {A comprehensive study on Structural Health Monitoring (SHM) of wind turbine blades by instrumenting tower using machine learning methods},
journal = {Renewable Energy},
volume = {199},
pages = {1568-1579},
year = {2022},
issn = {0960-1481},
doi = {10.1016/j.renene.2022.09.032},
author = {Meghdad Khazaee and Pierre Derian and Anthony Mouraud}}

@article{Pezeshki2023_intro,
    author = {Pezeshki, Hadi and Adeli, Hojjat and Pavlou, Dimitrios and Siriwardane, Sudath C.},
    title = {State of the art in structural health monitoring of offshore and marine structures},
    journal = {Maritime Engineering},
    volume = {176},
    number = {2},
    pages = {89-108},
    year = {2023},
    month = {01},
    issn = {1741-7597},
    doi = {10.1680/jmaen.2022.027}}

@article{xu2021design,
  title={Design and comparative analysis of alternative mooring systems for floating wind turbines in shallow water with emphasis on ultimate limit state design},
  author={Xu, Kun and Larsen, Kjell and Shao, Yanlin and Zhang, Min and Gao, Zhen and Moan, Torgeir},
  journal={Ocean Engineering},
  volume={219},
  pages={108377},
  year={2021},
  doi = {10.1016/j.oceaneng.2020.108377},
  publisher={Elsevier}
}

@article{chung2020detection,
  title={Detection of damaged mooring line based on deep neural networks},
  author={Chung, Minwoong and Kim, Seungjun and Lee, Kanghyeok and Shin, Do Hyoung},
  journal={Ocean Engineering},
  volume={209},
  pages={107522},
  year={2020},
  publisher={Elsevier},
  doi = {10.1016/j.oceaneng.2020.107522},
}

@article{mao2023LSTM-SVM,
  title={A new mooring failure detection approach based on hybrid {LSTM-SVM} model for semi-submersible platform},
  author={Mao, Yixuan and Zheng, Miaozi and Wang, Tianqi and Duan, Menglan},
  journal={Ocean Engineering},
  volume={275},
  pages={114161},
  year={2023},
  publisher={Elsevier}, 
  issn = {0029-8018},
  doi = {10.1016/j.oceaneng.2023.114161},
}

@article{CORADDU2024,
author = {Andrea Coraddu and Luca Oneto and Jake Walker and Katarzyna Patryniak and Arran Prothero and Maurizio Collu},
title = {Floating offshore wind turbine mooring line sections health status nowcasting: From supervised shallow to weakly supervised deep learning},
journal = {Mechanical Systems and Signal Processing},
volume = {216},
pages = {111446},
year = {2024},
issn = {0888-3270},
doi = {10.1016/j.ymssp.2024.111446},
}

@article{MAO2024,
title = {A novel mooring system anomaly detection framework for SEMI based on improved residual network with attention mechanism and feature fusion},
journal = {Reliability Engineering \& System Safety},
volume = {245},
pages = {109970},
year = {2024},
issn = {0951-8320},
doi = {10.1016/j.ress.2024.109970},
author = {Yixuan Mao and Xiaorong Li and Menglan Duan and Yongcun Feng and Jinjia Wang and Hongyuan Men and Heng Yang},
}

@article{sharma2024convolution,
  title={Condition monitoring of mooring systems for Floating Offshore Wind Turbines using Convolutional Neural Network framework coupled with Autoregressive coefficients},
  author={Sharma, Smriti and Nava, Vincenzo},
  journal={Ocean Engineering},
  volume={302},
  pages={117650},
  year={2024},
  doi = {10.1016/j.oceaneng.2024.117650},
  publisher={Elsevier}
  
}

@book{goodfellow2016deep,
  title={Deep Learning},
  author={Bengio, Yoshua and Goodfellow, Ian and Courville, Aaron and others},
  volume={1},
  year={2017},
  publisher={MIT press Cambridge, MA, USA, {I}SBN: 978-0-262-03561-3 },
}

@article{xu2022multisensory,
  title={Multisensory collaborative damage diagnosis of a 10 MW floating offshore wind turbine tendons using multi-scale convolutional neural network with attention mechanism},
  author={Xu, Zifei and Bashir, Musa and Yang, Yang and Wang, Xinyu and Wang, Jin and Ekere, Nduka and Li, Chun},
  journal={Renewable Energy},
  volume={199},
  pages={21--34},
  year={2022},
  publisher={Elsevier},
  doi = {10.1016/j.renene.2022.08.093},
}

@Article{Navamuel_wes2025,
AUTHOR = {Fernandez-Navamuel, A. and Gorostidi, N. and Pardo, D. and Nava, V. and Chatzi, E.},
TITLE = {Gaussian mixture autoencoder for uncertainty-aware damage identification in a floating offshore wind turbine},
JOURNAL = {Wind Energy Science},
VOLUME = {10},
YEAR = {2025},
NUMBER = {5},
PAGES = {857--885},
DOI = {10.5194/wes-10-857-2025}
}

@inproceedings{wang2019neural,
  title={Neural gaussian copula for variational autoencoder},
  author={Wang, Prince Zizhuang and Wang, William Yang},
  booktitle = {
    Proceedings of the 2019 Conference on Empirical Methods in Natural Language Processing and the 9th International Joint Conference on Natural Language Processing (EMNLP-IJCNLP)},
  year={2019},
  pages = {4333--4343},
  doi = {10.18653/v1/D19-1442},
}

@article{mo2025explainable,
  title={Explainable Neural-Networked Variational Inference: A New and Fast Paradigm with Automatic Differentiation for High-Dimensional Bayesian Inverse Problems},
  author={Mo, Jiang and Yan, Wang-Ji},
  journal={Reliability Engineering \& System Safety},
  pages={111337},
  year={2025},
  publisher={Elsevier},
  doi = { 10.1016/j.ress.2025.11133},
}

@book{nelsen2006introduction,
  title     = {An Introduction to Copulas},
  author    = {Nelsen, Roger B.},
  series    = {Springer Series in Statistics},
  edition   = {Second},
  year      = {2006},
  publisher = {Springer},
  address   = {New York},
  isbn      = {0-387-28659-4},
  note      = {{I}SBN 0-387-28659-4}
}

@inproceedings{nelsen2003properties,
  title={{Properties and Applications of Copulas: A brief Survey}},
  author={Nelsen, Roger B},
  booktitle={{Proceedings of the First Brazilian Conference on Statistical Modeling in Insurance and Finance}},
  pages={10--28},
  year={2003},
  organization={University Press USP Sao Paulo},
  url={https://api.semanticscholar.org/CorpusID:2508363}

}

@article{sklar1973random,
  title={Random variables, joint distribution functions, and copulas},
  author={Sklar, Abe},
  journal={Kybernetika},
  volume={9},
  number={6},
  pages={449--460},
  year={1973},
  url = {http://eudml.org/doc/28992},
  publisher={Institute of Information Theory and Automation AS CR}
}

@article{carbonera2024variational,
  title={Variational Autoencoders and Generative Adversarial Networks for Multivariate Scenario Generation},
  author={Carbonera, Michele and Ciavotta, Michele and Messina, Enza},
  journal={Data Science for Transportation},
  volume={6},
  number={3},
  pages ={23},
  year={2024},
  publisher={Springer},
  doi = {10.1007/s42421-024-00097-y},}

@book{mcneil2015quantitative,
  title     = {Quantitative Risk Management: Concepts, Techniques and Tools},
  author    = {McNeil, Alexander J. and Frey, R{\"u}diger and Embrechts, Paul},
  series    = {Princeton Series in Finance},
  edition   = {Revised},
  year      = {2015},
  publisher = {Princeton University Press},
  address   = {Princeton, NJ},
  isbn      = {978-0-691-16627-8},
  note      = {{I}SBN 978-0-691-16627-8}
}

@book{cherubini2011dynamic,
  title     = {Dynamic Copula Methods in Finance},
  author    = {Cherubini, Umberto and Mulinacci, Sabrina and Gobbi, Fabio and Romagnoli, Silvia},
  series    = {The Wiley Finance Series},
  year      = {2011},
  publisher = {John Wiley \& Sons},
  address   = {Hoboken, NJ},
  isbn      = {978-0-470-68307-1},
  note      = {{I}SBN 978-0-470-68307-1}
}

@article{coulling2013validation,
  title={Validation of a FAST semi-submersible floating wind turbine numerical model with DeepCwind test data},
  author={Coulling, Alexander J and Goupee, Andrew J and Robertson, Amy N and Jonkman, Jason M and Dagher, Habib J},
  journal={Journal of Renewable and Sustainable Energy},
  volume={5},
  number={2},
  pages={023116},
  year={2013},
  doi = {10.1063/1.4796197},
  publisher={American Institute of Physics}
}

@misc{suh2016gaussian,
      title={Gaussian Copula Variational Autoencoders for Mixed Data}, 
      author={Suwon Suh and Seungjin Choi},
      year={2016},
      primaryClass={stat.ML},
      url={https://arxiv.org/abs/1604.04960}, 
}

@misc{wu2023c,
      title={C$^2${VAE}: {G}aussian Copula-based {VAE} Differing Disentangled from Coupled Representations with Contrastive Posterior}, 
      author={Zhangkai Wu and Longbing Cao},
      year={2023},
      eprint={2309.13303},
      archivePrefix={arXiv},
      primaryClass={cs.LG},
      url={https://arxiv.org/abs/2309.13303}, 
}

@misc{zhong2021variational,
      title={Variational Autoencoders for Collaborative Filtering}, 
      author={Dawen Liang and Rahul G. Krishnan and Matthew D. Hoffman and Tony Jebara},
      year={2018},
      url={https://arxiv.org/abs/1802.05814}, 
}

@inproceedings{tagasovska2019copulas,
  author    = {Tagasovska, Natasa and Ackerer, Damien and Vatter, Thibault},
  title     = {Copulas as {H}igh-{D}imensional {G}enerative {M}odels: Vine Copula Autoencoders},
  booktitle = {Advances in Neural Information Processing Systems 32 (NeurIPS 2019)},
  year      = {2019},
  editor    = {H. Wallach and H. Larochelle and A. Beygelzimer and F. d\textquotesingle Alch\'{e}-Buc and E. Fox and R. Garnett},
  publisher = {Curran Associates, Inc.},
  address   = {Red Hook, NY},
  pages     = {6525--6537},
  url       = {https://proceedings.neurips.cc/paper/8880-copulas-as-high-dimensional-generative-models-vine-copula-autoencoders}
}

@book{sundar2017ocean,
  title={Ocean Wave Mechanics: Applications in Marine Structures},
  author={Sundar, V.},
  isbn={9781119241638},
  note = {{I}SBN 9781119241638 },
  series={Ane/Athena Books},
  url={https://books.google.es/books?id=gknhCgAAQBAJ},
  year={2017},
  publisher={Wiley}
}

@article{kamariotis2025Uncertainties,
    author = {Kamariotis, Antonios and Vlachas, Konstantinos and Ntertimanis, Vasileios and Koune, Ioannis and Cicirello, Alice and Chatzi, Eleni},
    title = {On the Consistent Classification and Treatment of Uncertainties in Structural Health Monitoring Applications},
    journal = {ASCE-ASME J Risk and Uncert in Engrg Sys Part B Mech Engrg},
    volume = {11},
    number = {1},
    pages = {011108},
    year = {2024},
    issn = {2332-9017},
    doi = {10.1115/1.4067140},
}

@incollection{joyce2003bayes,
  author       = {Joyce, James},
  title        = {{B}ayes' {T}heorem},
  booktitle    = {The Stanford Encyclopedia of Philosophy},
  editor       = {Edward N. Zalta},
  howpublished = {\url{https://plato.stanford.edu/entries/bayes-theorem/}},
  year         = {2003},
  edition      = {Fall 2003 Edition},
  publisher    = {Metaphysics Research Lab, Stanford University},
  address      = {Stanford, CA}
}

@article{jonkman2014hydrodyn,
  title={HydroDyn user’s guide and theory manual},
  author={Jonkman, Jason M and Robertson, AN and Hayman, Greg J},
  journal={National Renewable Energy Laboratory},
  year={2014},
  URL = {https://openfast.readthedocs.io/en/dev/source/user/hydrodyn/index.html}
}

@article{jonkman2015aerodyn,
  title={AeroDyn v15 user’s guide and theory manual},
  author={Jonkman, Jason M and Hayman, GJ and Jonkman, BJ and Damiani, RR and Murray, RE},
  journal={NREL Draft Report},
  volume={46},
  year={2015}
}

@article{yang2021investigation,
  title={Investigation on mooring breakage effects of a 5 {MW} barge-type floating offshore wind turbine using {F2A}},
  author={Yang, Yang and Bashir, Musa and Li, Chun and Wang, Jin},
  journal={Ocean Engineering},
  volume={233},
  pages={108887},
  year={2021},
  publisher={Elsevier},
  doi = {10.1016/j.oceaneng.2021.108887}}

@inproceedings{reig2024efficient,
  title={{An Efficient Frequency-Domain Based Methodology for the Preliminary Design of FOWT Substructures}},
  author={Reig, MA and Mendikoa, I and Petuya, V},
  booktitle={{Journal of Physics: Conference Series}},
  volume={2745},
  pages={012006},
  year={2024},
  organization={IOP Publishing},
  doi = {10.1088/1742-6596/2745/1/012006},
}

@inproceedings{rinker2020comparison,
  title={{Comparison of loads from HAWC2 and OpenFAST for the IEA Wind 15 MW Reference Wind Turbine}},
  author={Rinker, Jennifer and Gaertner, Evan and Zahle, Frederik and Skrzypi{\'n}ski, Witold and Abbas, Nikhar and Bredmose, Henrik and Barter, Garrett and Dykes, Katherine},
  booktitle={{Journal of Physics: Conference Series}},
  volume={1618},
  pages={052052},
  year={2020},
  organization={IOP Publishing},
  doi = {10.1088/1742-6596/1618/5/052052},
}

@inproceedings{spraul2017effect,
  title={{Effect of Marine Growth on Floating Wind Turbines
Mooring Lines Responses}},
  author={Spraul, Charles and Pham, Hong-Duc and Arnal, Vincent and Marine, Reynaud},
  booktitle={23e Congr{\`e}s Fran{\c{c}}ais de M{\'e}canique},
  pages={1--17},
  year={2017},
  URL = {https://api.semanticscholar.org/CorpusID:195655649}
}

@article{LIU2021391,
title = {Fault diagnosis of the 10MW Floating Offshore Wind Turbine Benchmark: A mixed model and signal-based approach},
journal = {Renewable Energy},
volume = {164},
pages = {391-406},
year = {2021},
issn = {0960-1481},
doi = {10.1016/j.renene.2020.06.130},
author = {Yichao Liu and Riccardo Ferrari and Ping Wu and Xiaoli Jiang and Sunwei Li and Jan-Willem van Wingerden},
}

@article{tran2015copula,
  title={Copula variational inference},
  author={Tran, Dustin and Blei, David and Airoldi, Edo M},
  journal={Advances in neural information processing systems},
  volume={28},
  year={2015},
  doi={https://doi.org/10.48550/arXiv.1506.03159},
}

@book{joe2014dependence,
  title     = {Dependence Modeling with Copulas},
  author    = {Joe, Harry},
  series    = {Monographs on Statistics and Applied Probability},
  volume    = {134},
  year      = {2014},
  publisher = {CRC Press},
  address   = {Boca Raton},
  isbn      = {978-1-4665-8322-1},
  note      = {{I}SBN 978-1-4665-8322-1}
}

@article{joe1993parametric,
  title={Parametric families of multivariate distributions with given margins},
  author={Joe, Harry},
  journal={Journal of multivariate analysis},
  volume={46},
  number={2},
  pages={262--282},
  year={1993},
  publisher={Elsevier},
  doi= {10.1006/jmva.1993.1061}}

@article{joe1996estimation,
  title={The estimation method of inference functions for margins for multivariate models},
  author={Joe, Harry and Xu, James Jianmeng},
  year={1996}
}

@article{gonzalez2025offshore,
  title={Offshore Wind and Energy Transition: Lessons Learned, Progress, and Trends},
  author={Gonz{\'a}lez, Mario OA and Jones, Dylan and Santiso, Andressa M and Akbari, Negar and Melo, David C and Nogueira, Luana P and Vasconcelos, Rafael M},
  journal={Annual Review of Environment and Resources},
  volume={50},  
  number = {1},
  pages = {409-432},
  year={2025},
  publisher={Annual Reviews},
  doi = {10.1146/annurev-environ-111523-102149}
}

@article{xia2022bayesian,
  title={Bayesian multiscale deep generative model for the solution of high-dimensional inverse problems},
  author={Xia, Yingzhi and Zabaras, Nicholas},
  journal={Journal of Computational Physics},
  volume={455},
  pages={111008},
  year={2022},
  publisher={Elsevier},
  doi={10.1016/j.jcp.2022.111008}
}

@article {graving2020vae,
	author = {Graving, Jacob M. and Couzin, Iain D.},
	title = {VAE-SNE: a deep generative model for simultaneous dimensionality reduction and clustering},
	elocation-id = {2020.07.17.207993},
	year = {2020},
	doi = {10.1101/2020.07.17.207993},
	publisher = {Cold Spring Harbor Laboratory},
	URL = {https://www.biorxiv.org/content/early/2020/07/17/2020.07.17.207993},
	journal = {bioRxiv}
}

@ARTICLE{bond2021deep,
  author={Bond-Taylor, Sam and Leach, Adam and Long, Yang and Willcocks, Chris G.},
  journal={IEEE Transactions on Pattern Analysis and Machine Intelligence}, 
  title={Deep Generative modelling: A comparative review of {VAEs}, {GANs}, normalizing flows, energy-based and autoregressive models}, 
  year={2022},
  volume={44},
  number={11},
  pages={7327-7347},
  doi={10.1109/TPAMI.2021.3116668}}

@article{portillo2020dimensionality,
doi = {10.3847/1538-3881/ab9644},
year = {2020},
publisher = {The American Astronomical Society},
volume = {160},
number = {1},
pages = {45},
author = {Portillo, Stephen K. N. and Parejko, John K. and Vergara, Jorge R. and Connolly, Andrew J.},
title = {Dimensionality Reduction of SDSS Spectra with Variational Autoencoders},
journal = {The Astronomical Journal},
}

@article{yong2022bayesian,
  title={Bayesian autoencoders with uncertainty quantification: Towards trustworthy anomaly detection},
  author={Yong, Bang Xiang and Brintrup, Alexandra},
  journal={Expert Systems with Applications},
  volume={209},
  pages={118196},
  year={2022},
  publisher={Elsevier},
  doi = {10.1016/j.eswa.2022.118196}
}

@inproceedings{prost2023inverse,
   title={Inverse problem regularization with hierarchical variational autoencoders},
   doi ={10.1109/ICCV51070.2023.02093},
   booktitle={2023 IEEE/CVF International Conference on Computer Vision (ICCV)},
   publisher={IEEE},
   author={Prost, Jean and Houdard, Antoine and Almansa, Andrés and Papadakis, Nicolas},
   year={2023},
   pages={22837–22848} }

@article{wu2022inverse,
author = {Wu, Hao and O’Malley, Daniel and Golden, John and Vesselinov, Velimir},
year = {2022},
month = {03},
title = {Inverse analysis with variational autoencoders: a comparison of shallow and deep networks},
volume = {3},
journal = {Journal of Machine Learning for Modeling and Computing},
publisher={Begel House Inc.},
doi = {10.1615/JMachLearnModelComput.2022042093} }

@inproceedings{nalisnick2016approximate,
  title={Approximate inference for deep latent {G}aussian mixtures},
  author={Nalisnick, Eric and Hertel, Lars and Smyth, Padhraic},
  booktitle={NIPS Workshop on Bayesian Deep Learning},
  volume={2},
  pages={131},
  year={2016},
  url={https://api.semanticscholar.org/CorpusID:11889762}}

@article{LIU201943,
title = {Variational inference with Gaussian mixture model and householder flow},
journal = {Neural Networks},
volume = {109},
pages = {43-55},
year = {2019},
issn = {0893-6080},
doi = {10.1016/j.neunet.2018.10.002},
author = {GuoJun Liu and Yang Liu and MaoZu Guo and Peng Li and MingYu Li},
}

@article{mcaliley2024stochastic,
  title={Stochastic inversion of geophysical data by a conditional variational autoencoder},
  author={McAliley, Wallace Anderson and Li, Yaoguo},
  journal={Geophysics},
  volume={89},
  number={1},
  pages={WA219--WA232},
  year={2024},
  publisher={Society of Exploration Geophysicists},
  doi = {10.1190/geo2023-0147.1}
}

@article{sahlstrom2023utilizing,
author = {Sahlstr\"{o}m, Teemu and Tarvainen, Tanja},
title = {Utilizing Variational Autoencoders in the Bayesian Inverse Problem of Photoacoustic Tomography},
journal = {SIAM Journal on Imaging Sciences},
volume = {16},
number = {1},
pages = {89-110},
year = {2023},
doi = {10.1137/22M1489897},
}

@article{coracca2023unsupervised,
  title={An unsupervised structural health monitoring framework based on Variational Autoencoders and Hidden {M}arkov Models},
  author={Cora{\c{c}}a, Eduardo M and Ferreira, Janito V and N{\'o}brega, Eur{\'\i}pedes GO},
  journal={Reliability Engineering \& System Safety},
  volume={231},
  pages={109025},
  year={2023},
  publisher={Elsevier},
  doi ={10.1016/j.ress.2022.109025}}

@ARTICLE{pollastro2023semi,
  author={Pollastro, Andrea and Testa, Giusiana and Bilotta, Antonio and Prevete, Roberto},
  journal={IEEE Access}, 
  title={Semi-Supervised Detection of Structural Damage Using Variational Autoencoder and a One-Class Support Vector Machine}, 
  year={2023},
  volume={11},
  pages={67098-67112},
  doi={10.1109/ACCESS.2023.3291674}}

@article{lin2024structural,
  title={Structural damage detection based on the correlation of variational autoencoder neural networks using limited sensors},
  author={Lin, Jun and Ma, Hongwei},
  journal={Sensors},
  volume={24},
  number={8},
  pages={2616},
  year={2024},
  publisher={MDPI},
  doi= {10.3390/s24082616}
}

@article{simpson2021machine,
  title={Machine learning approach to model order reduction of nonlinear systems via autoencoder and {LSTM} networks},
  author={Simpson, Thomas and Dervilis, Nikolaos and Chatzi, Eleni},
  journal={Journal of Engineering Mechanics},
  volume={147},
  number={10},
  pages={04021061},
  year={2021},
  publisher={American Society of Civil Engineers},
  url = {https://doi.org/10.48550/arXiv.2109.11213} }

@article{bacsa2025structural,
author = {Kiran Bacsa  and Wei Liu  and Imad Abdallah  and Eleni Chatzi },
title = {Structural Dynamics Feature Learning Using a Supervised Variational Autoencoder},
journal = {Journal of Engineering Mechanics},
volume = {151},
number = {2},
pages = {04024106},
year = {2025},
doi = {10.1061/JENMDT.EMENG-7635},
publisher={American Society of Civil Engineers},
}

@article{huang2025detecting,
author = {Shieh-Kung Huang and You-Jing Li and Yi-Xun Lin},
title ={Detecting and clustering structural damages using modal information and variational autoencoder with triple loss},
journal = {Structural Health Monitoring},
pages = {14759217241306720},
year={2025},
doi = {10.1177/14759217241306720},
publisher={SAGE Publications Sage UK: London, England}}

@article{Navamuel2022supervised,
  title={Supervised deep learning with finite element simulations for damage identification in bridges},
  author={Fernandez-Navamuel, Ana and Zamora-S{\'a}nchez, Diego and Omella, {\'A}ngel J and Pardo, David and Garcia-Sanchez, David and Magalh{\~a}es, Filipe},
  journal={Engineering Structures},
  volume={257},
  pages={114016},
  year={2022},
  publisher={Elsevier},
  doi = {10.1016/j.engstruct.2022.114016},
}

@article{Navamuel2022deepPCA,
  title={Deep learning enhanced principal component analysis for structural health monitoring},
  author={Fernandez-Navamuel, Ana and Magalhaes, Filipe and Zamora-S{\'a}nchez, Diego and Omella, Angel J and Garcia-Sanchez, David and Pardo, David},
  journal={Structural Health Monitoring},
  volume={21},
  number={4},
  pages={1710--1722},
  year={2022},
  publisher={SAGE Publications Sage UK: London, England}, 
  doi = {10.1177/14759217211041684},
}

@inproceedings{kviman2023cooperation,
  title={{Cooperation in the Latent Space: The Benefits of Adding Mixture Components in Variational Autoencoders}},
  author={Kviman, Oskar and Mol{\'e}n, Ricky and Hotti, Alexandra and Kurt, Semih and Elvira, V{\i}ctor and Lagergren, Jens},
  booktitle={International Conference on Machine Learning},
  pages={18008--18022},
  year={2023},
  publisher={PMLR},
  doi = {10.5555/3618408.3619151},}

@article{SampleAverageApprox2014guide,
  title={A guide to sample average approximation},
  author={Kim, Sujin and Pasupathy, Raghu and Henderson, Shane G},
  journal={Handbook of simulation optimization},
  pages={207--243},
  year={2014},
  publisher={Springer}, 
  url={https://api.semanticscholar.org/CorpusID:17021796}}

@InProceedings{bauer2019resampled,
  title = 	 {{Resampled Priors for Variational Autoencoders}},
  author =       {Bauer, Matthias and Mnih, Andriy},
  booktitle = 	 {{Proceedings of the Twenty-Second International Conference on Artificial Intelligence and Statistics}},
  pages = 	 {66--75},
  year = 	 {2019},
  editor = 	 {Chaudhuri, Kamalika and Sugiyama, Masashi},
  volume = 	 {89},
  series = 	 {Proceedings of Machine Learning Research},
  month = 	 {16--18 Apr},
  publisher =    {PMLR},
  url = 	 {https://proceedings.mlr.press/v89/bauer19a.html},
}

@article{robinson1991sampling,
    author = {Robinson, Enders and Clark, Dean},
    title = {Sampling and the Nyquist frequency},
    journal = {The Leading Edge},
    volume = {10},
    number = {3},
    pages = {51-53},
    year = {1991},
    month = {03},
    issn = {1070-485X},
    doi = {10.1190/1.1436812},}

@article{yu2008multimode,
  title={Multimode process monitoring with Bayesian inference-based finite Gaussian mixture models},
  author={Yu, Jie and Qin, S Joe},
  journal={AIChE Journal},
  volume={54},
  number={7},
  pages={1811--1829},
  year={2008},
  doi = {10.1002/aic.11515},
  publisher={Wiley Online Library}
}

@inproceedings{owen2000monte,
  title={{Monte Carlo, quasi-Monte carlo, and randomized quasi-Monte Carlo}},
  author={Owen, Art B},
  booktitle={{Monte-Carlo and Quasi-Monte Carlo Methods 1998: Proceedings of a Conference held at the Claremont Graduate University, Claremont, California, USA, June 22--26, 1998}},
  pages={86--97},
  year={2000},
  doi = {10.1007/978-3-642-59657-5_5. },
  organization={Springer}
}

@article{l2002recent,
  title={Recent advances in randomized quasi-Monte Carlo methods},
  author={L’Ecuyer, Pierre and Lemieux, Christiane},
  journal={Modeling uncertainty: An examination of stochastic theory, methods, and applications},
  pages={419--474},
  year={2002},
  publisher={Springer}
}

@article{frees1998understanding,
  title={Understanding Relationships Using Copulas},
  author={Frees, Edward W. and Valdez, Emiliano A.},
  journal={North American Actuarial Journal},
  volume={2},
  number={1},
  pages={1--25},
  year={1998},
  publisher={Taylor & Francis},
  doi = {10.1080/10920277.1998.10595667},
}

@article{patton2012,
  title={A review of copula models for economic time series},
  author={Patton, Andrew J.},
  journal={Journal of Multivariate Analysis},
  volume={110},
  pages={4--18},
  year={2012},
  publisher={Elsevier},
  doi = {10.1016/j.jmva.2012.02.021}}

@book{salvadori2007,
  title     = {Extremes in Nature: An Approach Using Copulas},
  author    = {Salvadori, Gianfausto and De Michele, Carlo and Kottegoda, Nath T. and Rosso, Renzo},
  series    = {Water Science and Technology Library},
  volume    = {56},
  year      = {2007},
  publisher = {Springer},
  address   = {Dordrecht},
  isbn      = {978-1-4020-4414-4},
  note      = {ISBN 978-1-4020-4414-4}
}

@article{Grana2017,
author = {Grana, Dario. and Fjeldstad, Torstein. and Omre, Henning.},
year = {2017},
month = {02},
pages = {},
title = {Bayesian {G}aussian mixture linear inversion for geophysical inverse problems},
volume = {49},
journal = {Mathematical Geosciences},
doi = {10.1007/s11004-016-9671-9}
}

@article{Figueiredo2019,
author = {Figueiredo, Leandro. and Grana, Dario. and Roisenberg, Mauro. and Rodrigues, Bruno.},
year = {2019},
month = {02},
pages = {1-53},
title = {Gaussian mixture mcmc method for linear seismic inversion},
volume = {84},
journal = {Geophysics},
doi = {10.1190/geo2018-0529.1}
}

@article{Vazquez2023, 
title={Joint stochastic simulation of petrophysical properties with elastic attributes based on parametric copula models}, 
volume={62}, 
DOI={10.22201/igeof.2954436xe.2023.62.2.1593}, 
number={2}, 
journal={Geofísica Internacional}, 
author = {Vázquez-Ramírez, Daniel. and Huong Le, Van. and Díaz-Viera, Martín. A. and del Valle-García, Raúl. and Erdely, Arturo.}, 
year={2023}, 
pages={487–506} 
}

@article{Vazquez2025,
title = {Joint geostatistical seismic inversion of elastic and petrophysical properties using stochastic cosimulation models based on parametric copulas},
journal = {Petroleum Science},
year = {2025},
issn = {1995-8226},
doi = {https://doi.org/10.1016/j.petsci.2025.10.029},
author = {D. Vázquez-Ramírez and M.A. Díaz-Viera and R. Valle-García},
}

@article{hernandez2024fast,
  title={Fast procedure to compute empirical and Bernstein copulas},
  author={Hern{\'a}ndez-Maldonado, Victor Miguel and Erdely, Arturo and D{\'\i}az-Viera, Mart{\'\i}n and Rios, Leonardo},
  journal={Applied Mathematics and Computation (preproof)},
  volume={477},
  pages={128827},
  year={2024},
  publisher={Elsevier}
}

\end{document}